\documentclass{article}     

\usepackage{amsmath, amsfonts,amsthm,amssymb}
\usepackage{algorithm}
\usepackage{algorithmic}

\usepackage[english]{babel}

\usepackage[left=1.75cm, top=2.7cm, bottom=2.7cm,right=1.75cm]{geometry}

\usepackage{latexsym,amssymb,amsfonts,graphicx}
\usepackage{amsmath,dsfont}
\usepackage{verbatim}
\usepackage{mathrsfs}
\usepackage{bm}
\usepackage{color}
\usepackage{multirow}

\usepackage{url}

\usepackage{bm}
\usepackage{natbib}

\def\pa{\partial}

\newcommand{\Px}{ \mathbb{P} }
\newcommand{\Qx}{ \mathbb{Q} }
\newcommand{\Ex}{ \mathbb{E} }
\newcommand{\Rx}{\mathbb{R}}

\newcommand{\gt}{\mathcal{G}_t}

\newcommand{\idc}{\mathbf{1}}

\newcommand{\Gx}{\mathbb{G}}
\newcommand{\Fx}{\mathbb{F} }

\newcommand{\G}{\mathcal{G}}

\newtheorem{theorem}{Theorem}[section]
\newtheorem{definition}{Definition}[section]

\newtheorem{proposition}[theorem]{Proposition}
\newtheorem{remark}[theorem]{Remark}
\newtheorem{lemma}[theorem]{Lemma}
\newtheorem{corollary}[theorem]{Corollary}

 \newcommand{\YGreen}{\color{YGreen}}
\definecolor{YGreen}{rgb}{0,0,0}

 \newcommand{\DGold}{\color{DGold}}
\definecolor{DGold}{rgb}{0,0,0}

     \newcommand{\DMagenta}{\color{Magenta}}
\definecolor{Magenta}{rgb}{0,0,0}
 
\definecolor{Orange}{rgb}{0,0,0}
\definecolor{DarkBlue}{rgb}{0,0,0}
\newcommand{\DBlue}{\color{DarkBlue}}
 \definecolor{Red}{rgb}{0.00, 0.00, 0.00}
    
    \definecolor{DRed}{rgb}{0.0, 0.0, 0.0}
    \newcommand{\DRed}{\color{DRed}}
    \definecolor{Blue}{rgb}{0.00, 0.00, 0.00}
    \newcommand{\Blue}{\color{Blue}}
        \definecolor{Green}{rgb}{0.0, 0.0, 0.0}
    
     \definecolor{Awesome}{rgb}{0.0, 0.0, 0.0}
    \newcommand{\Awesome}{\color{Awesome}}

    \definecolor{PaleGrey}{rgb}{0.0, 0.0, 0.0}

      \definecolor{Violet}{rgb}{0.0, 0, 0.0}
    \newcommand{\DViolet}{\color{Violet}}

   \definecolor{Cap}{rgb}{0, 0, 0}

\title{Pricing and {\DBlue Semimartingale} Representations of Vulnerable Contingent Claims in Regime-Switching Markets}
\author{Agostino Capponi\thanks{{School of Industrial Engineering, Purdue University, West  Lafayette, IN, 47907, USA ({\tt capponi@purdue.edu}).
}}
 \and Jos\'e E. Figueroa-L\'{o}pez\thanks{Department of Statistics, Purdue University, West Lafayette, IN, 47907,  USA ({\tt figueroa@purdue.edu}).}
 \and {Jeffrey Nisen}\thanks{Department of Statistics, Purdue University, West Lafayette, IN, 47907,  USA ({\tt jnisen@purdue.edu}).}}

\date{}

\begin{document}
\maketitle
\begin{abstract}
{\DBlue Using a suitable change of probability measure, we obtain a novel Poisson series representation for the arbitrage-free price process 
of vulnerable contingent claims in a regime-switching market driven by an underlying continuous-time Markov process. As a result of this representation, along with a short-time asymptotic expansion of the claim's price process, we} develop {\Blue an efficient} method for pricing claims {\DBlue whose payoffs} may depend on the full path of the underlying Markov chain. The proposed approach is applied to price not only simple European claims such as defaultable bonds, but also a new type of path-dependent {claims} that we term self-decomposable, as well as the important class of vulnerable call and put options on a stock. We provide a detailed error analysis and illustrate the accuracy and computational complexity of our {\DBlue method} on {\DBlue several} market traded instruments, such as defaultable bond prices, barrier options, and vulnerable call options. Using {\DBlue again} our Poisson series representation, we {\DBlue show} differentiability {\Awesome in time} of the pre-default price function {\Blue of European vulnerable claims}, {\DBlue which enables us to rigorously} deduce Feynman-Ka\v{c} representations for the pre-default pricing function and {\DBlue new semimartingale representations} for the price process of the vulnerable claim {\Blue under both risk-neutral and objective probability measures}.
\vspace{0.3 cm}



{\DMagenta \noindent{\textbf{Keywords and phrases}: Credit Risk, Regime-Switching Models, Option Pricing, {\DBlue Vulnerable Claims, Semimartingale Representations}.}}

\end{abstract}

\section{Introduction}
{\Awesome Regime-switching} models {are aimed} at capturing the {\DBlue appealing} idea that the macro-economy is {\DBlue subjected} to regular, yet unpredictable {\DBlue in time, regimes}, which in turn affect the {\Blue prices} of financial securities. For example, structural changes of macro-economic conditions, such as inflation and recession{\DBlue ,} may induce changes in the stock returns or in the term structure of interest rates and{\DBlue , }similarly, periods of high market turbulence and liquidity crunches may increase the default risk of financial institutions. This has been empirically verified in the stock market, {as stated by \cite{Ang2002b}, who {\DBlue proposed}} the existence of two regimes characterized by different levels of {volatility}. {Similar findings have also been documented in} the bond market {(see \cite{Ang2002a}, \cite{Ang2002c}, and \cite{Dai2007})}. Most recently, in the credit market, \cite{Longstaff} {\DBlue suggested} the existence of three regimes, associated with high, middle, and low default risk, via an empirical analysis of the corporate bond market over the course of the last 150 years.
These considerations have led many researchers to {use} {\Awesome regime-switching} models for asset pricing. In the context of options {\DBlue on stocks}, \cite{GuoPr} considered a market consisting of two regimes, and {provided} a semi-analytical formula for the option price, based on occupation time densities. \cite{BuffingtonEle} {generalized} the method to the case of multiple regimes, under the assumption that the generator of the Markov chain is time homogenous, and {derived} expressions for the price of European claims. {\Awesome \cite{GapJean} obtained closed form expressions for European claims, assuming geometric Brownian motion dynamics with regime-switching drifts, while \cite{Yao} {developed} a fixed point iteration scheme to recover {prices of European options} in a {\Awesome regime-switching} model, {assuming} time homogenous generators}.
{ \DRed {\Awesome Regime-switching} models for default-free interest rates derivatives have been studied by \cite{ellittwi}, \cite{elliotsiu}, and \cite{Kuena}.}  \cite{Norberg} {studied} no-arbitrage pricing of derivatives on the regime parameters with emphasis on computation.

Most of the literature has focused on {\DBlue default-free European style claims with ``simple" payoffs (i.e. payoffs depending only on the value of a random variable such as the terminal value of the underlying asset or the underlying source of randomness)} with the exception of a few works detailed next. {\Awesome \cite{GuoZhang} studied American-type default-free claims under {\Awesome regime-switching} models.  \cite{RogersGraz} provided a methodology to price barrier options with {\DBlue a} regime-switching dividend process.}
{\DRed In the context of credit risk, \cite{Kuen} {\Awesome considered} pricing of interest rate derivatives with a defaultable counterparty. \cite{Kuen08} {\Awesome analyzed} credit default swaps, and {\Awesome derived} a system of coupled partial differential equations yielding the term structure of default probabilities needed for pricing. \cite{Bielgame} and \cite{BielCrep08} {\Awesome studied} pricing and hedging of a defaultable game option under a Markov modulated default intensity framework.} {\Awesome \cite{Blan2004} gave} the hedging strategy of vulnerable claims using defaultable zero-coupon bonds and default-free assets, under the assumption that the default-free market is complete. \cite{BielJean05} {\Awesome applied} the theory of indifference pricing for pricing and replication of vulnerable claims within the reduced form framework. For an excellent survey on pricing and hedging aspects of vulnerable claims, the reader is referred to \cite{Biel04}.

{In this paper,} we consider a macro-economy with finitely many observable {\Blue economic} regimes, containing state information regarding the equity (drift and volatility), credit (hazard intensity and loss given default), and interest rate market {(short rate)}. {\DMagenta We consider} three liquidly traded securities, {namely, a money market account, a risky (default-free) stock, and a defaultable bond}. The dynamics of the securities {are assumed to} depend on the macro-economic regimes, which are modeled using a finite state continuous time Markov {process}.
We follow the reduced form approach to credit risk {\DBlue and} model the default event using a doubly stochastic framework.
{\DBlue Pricing of contingent claims} is still problematic in the context of regime-switching models, in part, due to the lack of easily computable {\DMagenta expressions} for option prices. {\DBlue Hence, explicit} characterizations of these quantities that allow {\DBlue their efficient evaluation} are of great need in the field. Our paper {\DMagenta contributes} to this fundamental problem by developing a novel {\DBlue approach} for {\DBlue pricing} vulnerable claims in a {\Awesome regime-switching} model consisting of an arbitrary {finite} number of regimes, whose dynamics is governed by a possibly time varying generator. {\DBlue The proposed method can also be applied to price a certain type of path dependent claims, referred hereafter as \textit{self-decomposable claims}, whose payoffs 
can} be decomposed in terms of payoffs of shorter maturity claims, of possibly different type  {(see Section \ref{sec:BscMthd} for the precise statement)}. {\DBlue Self-decomposable claims encompass not only basic} instruments, such as bonds, whose price may {also} be recovered from no-arbitrage arguments via the solution of a coupled system of ordinary differential equations (ODE{'}s), {\DBlue but also} more exotic {\DBlue path-dependent} instruments, where prices can only be recovered via Monte-Carlo methods. {\DBlue As an example, we apply} our algorithm {\DBlue to price a type of } barrier options on {\DBlue the volatility of the stock}, which turn out to be {self-decomposable} in terms of shorter maturity barrier {\DBlue options and} bond prices. {\DBlue Note that, in our macro-economic model, our regime driven parameters for the} short rate, volatility, and the default intensity may be seen as {\DBlue ``quantized"} proxies for {\DBlue the actual} interest rates, volatilities, and credit spread indices, and,  therefore, options on these {underlying financial measurements} provide means for the investor to hedge against interest rate risk, market and default risk, in different economic regimes.

{\DMagenta {\DBlue The key tool behind our results is a type of} Poisson series representation for {\DBlue the prices of} vulnerable claims {\DBlue obtained by applying} a novel change of {\DBlue probability} measure technique, {\DBlue which transforms the underlying Markov process into a ``homogenous" process with constant transition intensities}. 
{\DBlue This Poisson representation enables us to express the price of the claim as a series expansion in terms of} Laplace transforms of the symmetric Dirichlet distribution, {\DBlue which can be efficiently evaluated by Taylor approximations whenever the claim's maturity is small. Hence, a second important step in our method is to decompose the claim's price in terms of shorter dated claims. We provide a unified framework to quantify (and hence to control) the error due to the Taylor approximations, the error due to the discrete approximation of the time-varying generators, and even the error due to the truncation of the Poisson series representation.} 
 Our algorithm is computationally fast and, even for simple claims such as bonds, is able to achieve a high level of accuracy in the price, within a time complexity which compares favorably with standard ODE methods. As a far reaching application of our method, we {also} propose a new method to price vulnerable call/put options on the stock (where there is an additional risk factor represented by the Brownian motion), a subject which has received significant attention in the literature, as documented above.
%

Under mild assumptions on the Markov chain generator, we {\DBlue also apply our} Poisson series representation to {\Awesome give} a rigorous proof {\DBlue of} the differentiability of the pre-default price function. This, in turn, allows us {\Blue to provide} Feynman-Ka\v{c} and semimartingale representations for the pre-default price {\DRed process (given by Eq.~\eqref{Eq:DfltSecDf} below)}, and for the vulnerable claim price process {\DRed (given by Eq.~\eqref{Eq:DfltSecDfNet} below)}. This extends {\DBlue earlier results in the literature which, on one hand, had focused mostly on martingale representations and, on the other, typically presumed the time differentiability of the price function (e.g., \cite{Elliottkopp}, \cite{elliotcomm})}. Needless to say {\DBlue that our semimartingale representations} play a key role in the hedging of defaultable contingent claims, a problem which has attracted increasing attention since the start of the credit crisis, due to {\Awesome the high number of credit quality deteriorations, or even default events, experienced by financial institutions.} Although we do not {\DBlue explicitly} address the hedging problem in this paper, our formulas can be used within any hedging framework employing Markov-modulated dynamics. We refer the reader to \cite{BielRut03a}, \cite{BielRut03b}, \cite{Biel04},  \cite{Blan2004}, and \cite{BielCrep11} for hedging methodologies based on the mean-variance or BSDE approach.  {\Awesome {\DBlue The semimartingale representations} are also fundamental in portfolio optimization problems {\DBlue as demonstrated in \cite{CapFigDyn}. Another related work is \cite{Chuin11}, where} a martingale representation for a two-dimensional martingale generated by the Brownian motion and the Markov chain {\DBlue was provided}  with the objective {\DBlue of establishing} the existence of an admissible investment strategy in a default-free regime-switching market.}
}

{\DMagenta
The rest of the paper is organized as follows. Section \ref{sec:model} sets up the defaultable {\Awesome regime-switching} model. Section \ref{Sec:FKSerRepr} establishes the Poisson series representation of the vulnerable claim price function. Section \ref{sec:semi-mart} provides rigorous {\Blue semimartingale} representations for the price process of the {vulnerable claims}. Section \ref{sec:vulncl} {\DBlue presents our} novel efficient method for {pricing vulnerable claims} and {\DBlue provides} a rigorous error analysis.} Section \ref{sec:numerical} illustrates the accuracy and computational power of our algorithms on concrete choices of claims. Section \ref{sec:conclusion} concludes the paper. Proofs and numerical details are {\DBlue deferred} to the Appendix.

%

\section{{The defaultable regime-switching model}}\label{sec:model}
{\Blue As in \cite{CapFigDyn}}, we consider a {\Blue regime-switching market model with a defaultable security driven by a continuous-time Markov process $X$.} More specifically, our market consists of a risk-free asset, a risky (default-free) asset, and a defaultable $T$-claim {\Blue (such as a bond)} with respective price processes $\{B_{t}\}_{t\geq{}0}$, $\{S_{t}\}_{t\geq{}0}$, and {\Blue $\{\gamma(t,T)\}_{0\leq{}t\leq{}T}$} defined on a complete {\DMagenta filtered} probability space $(\Omega,\G,\Gx,\Px)$. Here, $\Px$ denotes the real world or historical probability measure and ${\Gx}:=(\gt)$ is an enlarged filtration given
by $\gt := \mathcal{F}_t \vee \mathcal{H}_t$, where $\Fx:=\{\mathcal{F}_{t}\}_{t\geq{}0}$ models the reference filtration and $\mathcal{H}_t = \sigma(H(u): u \leq t)$ is the filtration generated {by} an exogenous default process $H(t):= \idc_{\tau \leq t}$, after
completion and regularization on the right {(see \cite{Shreve} for details)}. {\Blue Throughout, $\tau$ represents the default time of the defaultable security of interest. We also adopt} the canonical construction of the default time $\tau$ in terms of a given hazard process {$\{h_t\}_{t\geq{}0}${,} so that
\begin{equation}
{\tau:=}\inf \left \{ t \in {\Blue \mathds{R}_+} : \int_0^t {h_u} du \geq {\YGreen \Theta} \right \},
\label{eq:taudef}
\end{equation}
where ${\YGreen \Theta}$ is an exponential random variable defined on the probability space $(\Omega,\G, \Px)$ and independent of $\Fx$. In that case, it follows that
\begin{equation}\label{MrtRprDftPrcP}
	{\xi^{\Px}_t} := H(t) - \int_0^t (1-H(u^{-})) h_u du
\end{equation}
is a ${\Gx}$-martingale under $\Px$ (see \cite{bielecki01}, Section 6.5).  {\Blue The intuition of (\ref{MrtRprDftPrcP}) is that one needs to compensate the single jump process for default prior to the occurrence of a default.} {\Blue Let us {\Blue also} recall that, under the previous setting, $\Fx$ satisfies the so-called \emph{martingale invariance property} with respect to $\Gx$, which states that every $\Fx$ martingale is also a $\Gx$ martingale (see Sections 8.3.1 and 8.6.1 in \cite{bielecki01}). The previous {\DBlue principle} is {\DBlue also} typically referred to as \emph{the H-hypothesis}}.

{\Blue As previously mentioned,} we place ourselves in a regime-switching market model. More specifically,
we define an $\Fx$-adapted continuous-time Markov process $\{X_{t}\}_{t\geq{}0}$ with finite state space $\{e_1, e_2, \ldots, e_N \}$, where hereafter $e_i = (0,...,1,...0)^{'} \in \mathds{R}^N$ and $'$ denotes the transpose. Throughout, {\Blue $p_{i,j}(t,s) := \Px(X_s = e_j | X_t = e_i)$} for {$t\leq{}s$} represents the transition probabilities of $X$ and $A(t):=[a_{i,j}(t)]_{i,j=1,2,\ldots,N}$ denotes the generator, {defined by}
\begin{equation}
	a_{i,j}(t) = \lim_{h \rightarrow 0} \frac{p_{i,j}(t,t+h)}{h},\quad (i\neq{}j), \qquad\quad a_{i,i}(t) := -\sum_{j \neq i} a_{i,j}(t).
\label{eq:chainrates}
\end{equation}
The following {\Blue semimartingale} representation is {\Blue well-known} (cf. \cite{elliottb}):
\begin{equation}
X_t = X_0 + \int_0^t A^{\prime}(s) X_s ds + {M^{\Px}(t)}{,}
\label{eq:MCsemim}
\end{equation}
where {$M^{\Px}(t)=(M^{\Px}_{1}(t),\dots,M_{N}^{\Px}(t))'$} is a $\mathds{R}^N$-valued $\Px$-martingale process. The following terminology will also be {\Blue frequently needed in the sequel}:
\begin{equation}\label{DfnCFunc}
	C_{t}:=\sum_{i=1}^{N} i {\bf 1}_{\{X_{t}=e_{i}\}}.
\end{equation}

We {\Blue now specify the three instruments in the market}, whose dynamics are driven by {\Blue $\{X_t\}_{t\geq{}0}$}. We have a locally \emph{risk-free asset} {\Blue $\{B_{t}\}_{t\geq{}0}$} with {\Blue Markov-modulated} dynamics
\begin{equation}\label{BPrDyn}
	{dB_t = r_t B_t d t,}
\end{equation}
where $r_t$ takes a constant value $r_{i}$ if the economy regime variable $X_{t}$ is at the $i^{th}$ state $e_{i}$. That is,
\[
	{\Blue r_{t}:=\left<r,X_t\right>=r_{_{C_{t}}}},
\]	
where {\Blue throughout} $\left<\cdot,\cdot\right>$ denotes the standard inner product in $\mathbb{R}^{N}$ and $r=(r_1,r_2,\ldots,r_N)'$ {\Blue is a vector of} positive constants, {\Blue which represent the possible risk-free short-rates in the market}. The risky (default-free) asset {\Blue $\{S_{t}\}_{t\geq{}0}$} follows the dynamics
\begin{equation}\label{SPNDD}
dS_t = \mu_t S_t dt + \sigma_t S_t dW_t, \qquad S_0 = s,
\end{equation}
where {\Blue $\{W_{t}\}_{t\geq{}0}$} is an $\Fx$-adapted {\Blue standard Brownian motion} independent of $\{X_{t}\}$ {\Blue under the measure $\Px$}, and
\begin{equation}
{\Blue \mu_t:=  \left<\mu, X_t\right>={\DRed \mu_{_{C_{t}}}}, \qquad \sigma_t := \left<\sigma,X_t\right>=\sigma_{_{C_{t}}}}
\end{equation}
for constant vectors $\mu = (\mu_1,\mu_2,\ldots,\mu_N)'$ and $\sigma = (\sigma_1,\sigma_2,\ldots,\sigma_N)'${,} representing the respective appreciation rates and volatilities that the risky asset can take depending on the different economic regimes.

{\Blue The pricing methods we introduce in this paper can be applied to a wide range of defaultable securities such as corporate bonds, vulnerable claims, recovery payments at default, or even vulnerable call/put options on the risky stock $\{S_{t}\}_{t\geq{}0}$. As a way to fix ideas, we will focus on a defaultable security, whose recovery payment at default is determined by the \emph{recovery of market value} (RM) assumption (see, e.g.,  \cite[Section 9.4]{Embrechts} or \cite[Section 8.3]{bielecki01}).  More precisely, this security delivers an $\mathcal{F}_{T}$-measurable promised payment $\mathcal{X}$ at time $T$ if no default has occured by that time (i.e. $\tau>T$) and, otherwise, delivers the recovery payment $(1-L_{\tau}) \gamma(\tau^{-};T)$ at the default time $\tau\leq{}T$, where $\{\gamma(t;T)\}_{t\geq{}0}$ is the \emph{pre-default value process} of the security and $\{L_{t}\}_{t\geq{}0}$ is an $\Fx$-adapted process, which represents the percentage loss given default. Then, the following fundamental risk-neutral valuation formula due to \cite{ds} (see also Proposition 8.3.3 in \cite{bielecki01}) is well-known: 
\begin{equation}\label{Eq:DfltSecDf}
	\gamma(t;T) = \Ex^{\Qx} \left[ \mathcal{X} \, e^{-\int_t^T (r_s + h_{s} L_{s}) ds} \bigg| \mathcal{F}_t \right],
\end{equation}
where $\Qx$ is the equivalent martingale measure used in pricing. Furthermore, denoting $\{\Gamma(t;T)\}_{t\geq{}0}$ the value of the previously described defaultable security, it follows that
\begin{equation}\label{Eq:DfltSecDfNet}
	\Gamma(t;T) = {\bf 1}_{\tau>t}\gamma(t;T)= {\bf 1}_{\tau>t}\Ex^{\Qx} \left[ \mathcal{X} \, e^{-\int_t^T (r_s + h_{s} L_{s}) ds} \bigg| \mathcal{F}_t \right].
\end{equation}
 An important example of the defaultable security (\ref{Eq:DfltSecDfNet}) is a defaultable bond given by the price process
\begin{equation}
p(t,T) = {{\bf 1}_{\{\tau>t\}} \Ex^{\Qx} \left[e^{-\int_t^T (r_s + {h_s} {L_s})ds}  \bigg| \mathcal{F}_t \right]}.
\label{eq:bondpr}
\end{equation}

Throughout, we assume that $\Qx$ is such that $W$ is still a standard Brownian motion and $X$ is an independent continuous-time Markov process, under $\Qx$, with possibly different generator $A^{\Qx}(t):=[a_{i,j}^{\Qx}(t)]_{i,j=1,2,\ldots,N}$.
The existence of the measure $\Qx$ follows from the theory of change of measures for continuous-time Markov chains (see Section 11.2 in \cite{bielecki01}). Such density transformation theorems for Markov chains will play a key role in our pricing approach below. Concretely, for any $i\neq{}j$ and some bounded measurable functions  $\kappa_{i,j}:\mathbb{R}_+\to (-1,\infty)$, let
\begin{equation}\label{DfnRNGen}
	a_{i,j}^{\Qx}(t):=a_{i,j}(t) (1+\kappa_{i,j}(t)),
\end{equation}
and, for $i=j$, let
\[
	a_{i,i}^{\Qx}(t):=-\sum_{k=1,k\neq{}i}^{N}a_{i,k}^{\Qx}(t).
\]
We also fix $\kappa_{i,i}(t)=0$ for $i=1,\dots,N$.
Now, consider the processes
\begin{equation}\label{CrsMrt}
	M_{t}^{i,j}:=H^{i,j}_{t}-\int_{0}^{t} a_{i,j}(u)
	H^{i}_{u}du,
\end{equation}
where
\begin{equation}\label{JmpTrnPrc1}
	H_{t}^{i}:= \idc_{\{X_{t}=e_i\}}, \quad \text{and}\quad
	 H_{t}^{i,j}:=\sum_{0<u\leq{}t}\idc_{\{X_{u^{-}}=e_{i}\}}\idc_{\{X_{u}=e_{j}\}}, \quad (i\neq{}j).
\end{equation}
The process $\{M^{i,j}_{t}\}_{t\geq{}0}$ is known to be  an ${\Fx}$-martingale for any $i\neq{}j$ (see Lemma 11.2.3 in \cite{bielecki01}) and, since the $H$-hypothesis holds in our default framework, they are also ${\Gx}$-martingales.
Then, we have the following result (see Proposition 11.2.3 in \cite{bielecki01}):
\begin{lemma}
The probability measure $\Qx$ {on {$\Gx=(\mathcal{G}_{t})_{t}$}} with {Radon-Nikodym} density $\{\eta_{t}\}_{t\geq{}0}$ given by
\begin{equation}\label{DfnDnsty}
	\eta_{t}=1+\int_{(0,t]} \sum_{i,j=1}^{N} \eta_{u^{-}} \kappa_{i,j}(u) d M_{u}^{i,j},\qquad t\geq{}0,
\end{equation}
is such that $X$ is a Markov process under $\Qx$ with generator $[a_{i,j}^{\Qx}(t)]_{i,j=1,2,\ldots,N}$.
\end{lemma}
Without loss of generality, one can take the measure $\Qx$ in the previous lemma to be such that $W$ is still a Wiener process independent of $X$ under $\Qx$.  Analogously to (\ref{eq:MCsemim}), the process
\begin{equation}
M^{\Qx}(t):= X_t - X_0 - \int_0^t A^{\Qx}(s)' X_s ds,
\label{eq:MCsemimb}
\end{equation}
is a $\mathds{R}^N$-valued martingale under $\Qx$.  Comparing (\ref{eq:MCsemimb}) to (\ref{eq:MCsemim}), note that
\begin{equation}\label{RMM0}
	M^{\mathbb{Q}}(t)= M^{\mathbb{P}}(t)+\int_{0}^{t} ({A(s)'}-A^{\mathbb{Q}}(s)')X_{s}ds.
\end{equation}

To be consistent with our regime-switching market model, we assume {\DBlue that the} default intensity $h$ and loss rates $L$ are also driven by the underlying Markov process $X$; i.e.,
\begin{equation}
h_t:=  \left<h, X_t\right>=h_{_{C_{t}}}, \qquad\text{ and }\qquad L_t:= \left< L,X_t\right>=L_{_{C_{t}}},
\end{equation}
for constant vectors $h = (h_1,\ldots,h_N)'$ and $L = (L_1,\ldots,L_N)'${\DRed, where $h_i>0$, and  $0 \leq L_i \leq 1$}. We emphasize that the distribution of the hazard rate process $h_{t}=\left<h,X_{t}\right>$ under the risk-neutral measure is different from that under the historical measure, {\DBlue which allows to incorporate a default risk premium into our framework.}
{\Awesome 
}

The following result states that the process $\{\xi^{\Px}_t\}_{t\geq{}0}$ introduced in (\ref{MrtRprDftPrcP}) is also a {$\Qx$-martingale}. However, in order to indicate in the sequel when certain dynamics are being taken under $\Qx$ or under  $\Px$, we will introduce a new notation $\{\xi^{\Qx}_t\}_{t\geq{}0}$. We should {\Awesome keep} in mind throughout that $\xi^\Qx=\xi^\Px$. The proof of this {\Awesome result} is presented in Appendix \ref{NmrMthdApdx}.
\begin{lemma}\label{LmNdFD}
	The process
	\begin{equation}\label{MrtRprDftPrcQ}
		{\xi^{\Qx}_t} :=H(t) - \int_0^t (1-H(u^{-})) {h_u} du
	\end{equation}
	is also a $({\Gx},\Qx)-$(local) martingale.
\end{lemma}
\section{Poisson series representations for vulnerable claims}\label{Sec:FKSerRepr}

{\DRed In this section, we introduce the main tool upon which the construction of our novel algorithms as well as the semimartingales and Feynman-Ka\v{c} representations of the price processes are based on}. We focus on the case when the promised payoff $\mathcal{X}$ in (\ref{Eq:DfltSecDfNet})} is of the form $\Xi(X_{T})$ for a deterministic function $\Xi:\{e_1,e_2,\dots,e_N\}\to{}\Rx$. In this case, due to the Markov property, the pre-default value process (\ref{Eq:DfltSecDf}) can be written as
\[
	\gamma(t;T)=\Psi_{_{C_{t}}}[\Xi](t;T),
\]
where, for each economic regime $i\in\{1,\dots,N\}$,
\begin{equation}\label{BndPrc3}
	\Psi_{i}[\Xi](t;T):=\Ex^{\Qx}\left[\left. \Xi(X_{T})e^{-\int_{t}^{T}(r_{s} + h_{s}L_{s})ds} \right| X_{t} = e_{i} \right].
\end{equation}

The key idea of our {approach} {to compute (\ref{BndPrc3}) lies in} changing the risk-neutral probability measure $\Qx$ into an equivalent measure $\widetilde{\Qx}$ such that $\{X_{t}\}_{t\leq{}T}$ {\Blue becomes} a homogeneous Markov process under $\widetilde{\Qx}$. Such a probability measure $\widetilde{\Qx}$ exists whenever {\Blue $a_{i,j}^{\Qx}(t)$ is strictly positive for all $t>0$ and $i\neq{}j$}. Concretely, define functions $\tilde{\kappa}_{i,j}:[0,\infty)\to(-1,\infty)$ such that
	\[
		{\Blue \frac{1}{N-1}}=a_{i,j}^{\Qx}(t) (1+\tilde\kappa_{i,j}(t)),\quad\text{ for any } i\neq{}j, \quad\text{and}\quad
		\tilde{\kappa}_{i,i}=0.
	\]
	We also let $\tilde{a}_{i,j}:={\Blue 1/(N-1)}$ for any $i\neq{}j$ and $\tilde{a}_{i,i}={\Blue -1}$ for any $i=1,\dots,N$, so that $\widetilde{A}:=\left[\tilde{a}_{i,j}\right]_{i,j=1,\dots,N}$ is a valid generator of a homogeneous Markov process with transition times determined by a homogeneous Poisson process $\{M_{t}\}_{t\geq{}0}$ with intensity $1$ and an embedded Markov chain $\{\widetilde{X}_{i}\}_{i\geq{}1}$ with transition probabilities $p_{i,j}:=1/(N-1)$ for $i\neq{}j$.  Now, let us define a probability measure $\widetilde{\Qx}$ with {Radon-Nikodym} density $\{\tilde\eta_{t}\}_{t\geq{}0}$ given by
\begin{equation}\label{DfnDnstyHom}
	\tilde\eta_{t}=1+\int_{(0,t]} \sum_{i,j=1}^{N} \tilde\eta_{u^{-}} \tilde\kappa_{i,j}(u) d \widetilde{M}_{u}^{i,j},
\end{equation}
where
\(
	\widetilde{M}_{t}^{i,j}:= H_{t}^{i,j}-\int_{0}^{t} a_{i,j}^{\Qx}(u)H^{i}_{u}du,
\)
and we used notation (\ref{JmpTrnPrc1}). By virtue of Proposition 11.2.3 in \cite{bielecki01},  $\{X_{t}\}_{t\geq{}0}$ is a continuous Markov process with generator $\widetilde{A}$ under {\Blue $\widetilde{\Qx}$}. The representation {\DBlue below then follows:}
{\DRed \begin{theorem}\label{Lm:DfPsi}
	Suppose that, for any $i\neq{}j$, the function $a_{i,j}^{\Qx}$ defined in (\ref{DfnRNGen}) is such that
	\begin{equation}\label{NdCndDf1}
		0<\inf_{s\in[0,T]}|a_{i,j}^{\Qx}(s)|\leq{}
		\sup_{s\in[0,T]}|a_{i,j}^{\Qx}(s)|<\infty.
	\end{equation}
	Then,
	\begin{equation}\label{OptPrc2}
		\Psi_{i}[\Xi](t;T)=\Ex^{\widetilde{\Qx}}\left[\left. \Xi(X_{T})e^{-\int_t^T {\tilde{r}(s)'} {X}_{s} ds
		- \sum_{\{s\in(t,T]:\Delta X_{s}\neq{}0\}}  X_{s^{-}}'\widetilde{K} (s) X_{s}}\right| X_{t}=e_{i}\right],
	\end{equation}
	where  $\widetilde{K}(t):=[\widetilde{K}_{i,j}(t)]_{i,j}$ and $\tilde{r}(t):=(\tilde{r}_{1}(t),\dots,\tilde{r}_{N}(t))'$ are defined as
	\begin{equation}\label{Eq:DfnTildeKTilder}
		\widetilde{K}_{i,j}(t):=-\log\left((N-1)a_{i,j}^{\Qx}(t)\right){{\bf 1}_{i\neq j}},\qquad
		\tilde{r}_{i}(t):=r_{i}+h_{i}L_{i}- 1- a_{i,i}^{\Qx}(t).
	\end{equation}
 \end{theorem}}	
As explained before, $\widetilde{\Qx}$ has {the virtue} that the generator of $\{X_{t}\}_{t\geq{}0}$ under $\widetilde{\Qx}$ is given by $\tilde{a}_{i,j}=1/(N-1)$ ($i\neq{}j$) and $\tilde{a}_{i,i}=-1$. Thus, by conditioning on the number of transitions in $(t,T]$ (which is {\DBlue necessarily} Poisson distributed {with unit intensity}), we get the following series representation:
\begin{equation}\label{CmpPsii}
	\Psi_{i}[\Xi](t;T)=\sum_{m=0}^{\infty} e^{-(T-t)}\frac{(T-t)^{m}}{m!}\Phi_{i,m}[\Xi](T-t;T),
\end{equation}
where, denoting $\Ex^{\widetilde{\Qx}}_{i}\left[\cdot\right]=\Ex^{\widetilde{\Qx}}\left[\left.\cdot\right|{\Blue \widetilde{X}_{0}=e_{i}}\right]$,
\begin{align}\label{AuxDfn1b}
	{\Blue \Phi_{i,m}[\Xi](\zeta;T):=\Ex^{\widetilde{\Qx}}_{i}\left[ \Xi(\widetilde{X}_{m})\exp\left\{
	-\sum_{n=0}^{m} \int_{\zeta U_{(n)}}^{\zeta U_{(n+1)}}{\tilde{r}(T-\zeta+s)'} \widetilde{X}_{n} ds -\sum_{n=1}^{m}\widetilde{X}_{n-1}'\widetilde{K} (T-\zeta+\zeta U_{(n)}) \widetilde{X}_{n}\right\}\right].}
\end{align}
{\DBlue Here,} $\{\widetilde{X}_{i}\}$ is the embedded Markov chain of $\{X_{t}\}_{t\geq{}0}$ and $U_{(1)}<U_{(2)}<\dots<U_{(m)}$ are the ordered statistics of $m$ i.i.d. uniform $[0,1]$ variables independent of $\{\widetilde{X}_{i}\}$, fixing $U_{(0)}:=0$ and $U_{(m+1)}:=1$\footnote{{\Blue Note that, strictly speaking, the distribution of the variables $0=U_{(0)}<U_{(1)}<\dots<U_{(m)}<U_{(m+1)}=1$ depends on $m$ and, hence, we should probably denote them as $\{U^{(m)}_{(i)}\}_{i=1}^{m}$, but for simplicity, we will omit the superscript $(m)$.}}. {\Blue Thus, for instance, we have:}
\begin{equation}\label{Eq:1st2ndCoeff}
	{\Blue \Phi_{i,0}[\Xi](\zeta;T)=\Xi(e_{i}),\quad
	 \Phi_{i,1}[\Xi](\zeta;T)=\frac{\zeta^{-1}}{N-1}\sum_{j\neq{}i}\Xi(e_{j})\int_{T-\zeta}^{T}e^{-\int_{0}^{v}\tilde{r}_{i}(w)dw-\int_{v}^{T} \tilde{r}_{j}(w)dw-\widetilde{K}_{i,j}(v)} dv.}
\end{equation}

{\Blue {\DViolet The} series representation (\ref{CmpPsii}) is one of the {\DBlue fundamental} tools in the forthcoming results. 
We shall apply it to {\Blue find conditions} for the pre-default price functions (\ref{BndPrc3}) to be differentiable, which in turn {\Blue will lead to rigorous} Feynman-Ka\v{c} and semimartingale representations for the pre-default {\Awesome price functions (\ref{Eq:DfltSecDf})} and the vulnerable claim price process (\ref{Eq:DfltSecDfNet}), respectively (see Section \ref{sec:semi-mart} below). Furthermore, we shall also apply the above Poisson series representation in developing our pricing methods for {\Blue vulnerable {\YGreen contingent} claims} (see Section \ref{sec:vulncl} below).} {{\DBlue In what follows,} the dependence on $T$ will {\DBlue sometimes} be omitted from $\Phi$ in order to lighten {\YGreen the} notation}.

\section{Semimartingale representations of vulnerable claims} \label{sec:semi-mart}
In this section, we {\DViolet rigorously} deduce Feynman-Ka\v{c} martingale representations for the pre-default {price} process (\ref{BndPrc3}), and {\DBlue semimartingale} representations for the {price} process of the vulnerable claim given by Eq.~(\ref{Eq:DfltSecDfNet}). {The latter are given both under the risk-neutral probability measure $\Qx$ {\DBlue and} under the historical probability measure $\Px$.} {\Blue The representations} will be expressed in terms of the martingales (\ref{MrtRprDftPrcP}) and (\ref{eq:MCsemim}) (respectively, (\ref{eq:MCsemimb}) and (\ref{MrtRprDftPrcQ})). 
The semimartingale representations obtained here {\Blue are vital} in many fundamental problems of finance including hedging and portfolio optimization. {\DMagenta Hedging with credit risky} securities has attracted a lot of attention in recent years (see, e.g., \cite{BielRut03a}, \cite{BielRut03b}, \cite{Biel04}, and \cite{Blan2004}), while, combined with our portfolio optimization framework {\Blue developed} in \cite{CapFigDyn}, our results here can also be applied to portfolio optimization problems with more {\DMagenta general vulnerable claims} of the form (\ref{Eq:DfltSecDfNet}).

The {\Blue representations} obtained in this section are based on the following result, which in turn follows from the Poisson series representation (\ref{CmpPsii}) introduced in the previous {section}. Its proof is presented in Appendix \ref{NmrMthdApdx}.}
\begin{lemma}\label{Cor:DiffPrice}
In addition to the conditions in (\ref{NdCndDf1}), suppose that the functions $a_{i,j}^{\Qx}$ are continuously differentiable in $(0,T)$ {\DBlue and that}
	\begin{equation}\label{NdCndDf1b}
		 \sup_{s\in[0,T]}\left|\frac{d a_{i,j}^{\Qx}(s)}{ds}\right|<\infty, \qquad \text{for all }i\neq{}j.
	\end{equation}
Then, the pre-default price function $\Psi_i(t):=\Psi_i[\Xi](t;T)$ defined in (\ref{BndPrc3}) is continuously differentiable for any $t\in(0,T)$.
\end{lemma}

\subsection{Markov chain dependent claims}
{\DMagenta
We {\Blue proceed to} {\DBlue establish a system of} {\Blue ordinary} differential equations {\Awesome for the pre-default price functions (\ref{Eq:DfltSecDf})} and the {\Blue semimartingale} representation {\DBlue for} the price process {\Blue (\ref{Eq:DfltSecDfNet}) in the case of a claim $\mathcal{X}$ depending} only on the underlying Markov chain. {\Blue An important example of a Markov chain dependent claim is the defaultable bond price, which amounts to taking {\Blue $\Xi(e_{i}) = 1$ for all $i=1,\dots,N$}.} Similar {\Blue Feynman-Ka\v{c}} representations have been given in \cite{Elliottkopp}, Corollary 9.8.4. However,
these are given under the assumption that the price of the regime conditioned claim is a time differentiable function. 
Here, we fully justify the differentiability assumption using our {\DRed Theorem
\ref{Lm:DfPsi} and Lemma \ref{Cor:DiffPrice}} {\DBlue and, thus,} make the analysis fully rigorous. {\Blue To the best of our knowledge, the semimartingales representations are new and, as previously indicated, are crucial in, e.g., hedging and portfolio optimization problems.}}

{\Blue The following result, whose proof is presented in Appendix \ref{NmrMthdApdx},  is the main result of this section}.
\begin{theorem}\label{Th:FKRNDyn}
Suppose that the conditions (\ref{NdCndDf1}) and (\ref{NdCndDf1b}) are satisfied.
Then, the pre-default price function $\Psi_i(t):=\Psi_i[\Xi](t;T)$ is differentiable for any $t\in(0,T)$. Furthermore, the {\DMagenta following statements hold}:
\begin{enumerate}
	\item[{\bf (1)}] The vector of pre-default prices {\Blue $\Psi(t) = \left(\Psi_1(t),\ldots,\Psi_N(t) \right)'$} {\DMagenta satisfies} the following coupled system of {\DMagenta backward {Ordinary} Differential Equations (ODE)}:
\begin{align}
 \nonumber \frac{d\Psi_i(t)}{dt} &= {(r_i + h_i L_i - a_{i,i}^{\Qx}(t)) \Psi_i(t) - \sum_{j \neq i} a_{i,j}^{\Qx}(t) \Psi_j(t)}, {\Blue \qquad 0 < t < T,} \\
\Psi_i(T) &= \Xi(e_{i}), \qquad i = 1,\ldots,N.
\label{eq:bondspdeGen}
\end{align}
\item[{\bf (2)}] The prices process (\ref{Eq:DfltSecDfNet}) of the vulnerable claim with payoff $\Xi(X_{T})$ {\Awesome has the following {\DBlue semimartingale} representation:}
\begin{align}\label{DRNM0}
d\,\Gamma(t,T)&=\Gamma(t^{-};T)\left( [r_t + h_t (L_t-1)] dt + \frac{{\left<{\Psi(t)}, dM^{\Qx}(t)\right>}}{{\left<\Psi(t), X_{t^{-}}\right>}} - d\xi_t^{\Qx}\right),\quad 0\leq{}t<T, \quad {\Awesome t < \tau,}
\end{align}
under the risk-neutral probability measure $\Qx${\DBlue .}
\item[{\bf (3)}] {\Blue The prices process (\ref{Eq:DfltSecDfNet}) of the vulnerable claim with payoff $\Xi(X_{T})$ {\Awesome has the following {\DBlue semimartingale} representation:}
\begin{equation}\label{DRWP0}
d\,\Gamma(t,T)=\Gamma (t^{-},T){\DMagenta \left( \left[r_t + h_t (L_t-1)+{D(t)} \right] dt + \frac{\left<{\Psi(t)}, dM^{\Px}(t)\right>}{\left<\Psi(t), X_{t^{-}}\right>}-d\xi^{\Px}_{t} \right)}, \quad 0\leq{}t<T,\quad  {\Awesome t < \tau,}
\end{equation}
under the historical probability measure $\Px$, where $D(t):=\left< (D_{1}(t),\dots,D_{N}(t))',X_{t}\right>$ with}
\begin{equation}\label{NDfnDf}
	{\Blue D_{i}(t):=\sum_{j=1}^{N} (a_{i,j}(t)-a_{i,j}^{\Qx}(t))\frac{\Psi_{j}(t)}{\Psi_{i}(t)}
	= \sum_{j\neq i} (a_{i,j}(t)-a_{i,j}^{\Qx}(t))\left(\frac{\Psi_{j}(t)}{\Psi_{i}(t)}-1\right).}
\end{equation}
\end{enumerate}
\end{theorem}

\begin{remark}
As a direct consequence of the previous result, in the time-invariant case ({i.e.,} $a_{i,j}^{\Qx}(t)\equiv a_{i,j}^{\Qx}$), the solution of (\ref{eq:bondspdeGen}) can be expressed in ``closed-form" as
\begin{equation}
{\psi(t;T) = e^{-(T-t) F_{\psi}}\, {\bf \Xi}},
\label{eq:psiexpr}
\end{equation}
where ${\bf \Xi}=(\Xi(e_{1}),\dots,\Xi(e_{N}))'$ and the components of the matrix $F_{\psi}$ are
\[
	[F_{\psi}]_{i,i} = r_i + h_i L_i - {a^{\Qx}_{i,i}},\quad\text{ and }\quad [F_{\psi}]_{i,j} = -a_{i,j}^{\Qx}, \quad i\neq{}j.
\]
{\DRed
}
\end{remark}

\subsection{Markov chain and {\DBlue diffusion} dependent claims}
{\DMagenta
In this section, we provide the {\Awesome Feynman-Ka\v{c} {\DBlue type} partial differential equations} {\DBlue and the semimartingale} {\Blue representations} of the price process for claims {\Blue whose payoffs} depend both on the Brownian paths as well as on the underlying Markov chain. {\Blue For non-vulnerable claims}, \cite{elliotcomm} and \cite{Yao} {\Blue obtained similar Feynman-Ka\v{c} representations}. Here, the analysis is extended to deal with vulnerable claims.

We follow \cite{elliottet} {\DBlue and} choose the {\Blue the risk-neutral measure $\Qx$} {\DBlue such that}
\begin{equation}
\frac{d\Qx}{d\Px} \bigg|_{\mathcal{F}_t} = \exp \left \{\int_0^t \left( \frac{r_s - \mu_s}{\sigma_s} \right) dW_s - \frac{1}{2} \int_0^t \left(\frac{r_s - \mu_s}{\sigma_s} \right)^2 ds \right \}.
\label{eq:radonnik}
\end{equation}
{\Blue In particular}, the ($\mathcal{F}_t, \Qx$) Brownian motion is given by
\begin{equation}
W_t^{\Qx} = W_t + \int_0^t \left( \frac{r_s - \mu_s}{\sigma_s} \right) ds.
\label{eq:Qbrown}
\end{equation}
{\Blue Note that, without loss of generality, we can take $\Qx$ so that under it, $\{X_{t}\}_{t\geq{}0}$ is still a continuous-time Markov process with generator $A^{\Qx}$, independent of $\{W^{\Qx}_{t}\}_{t\geq{}0}$.}

Next, we consider a pre-default {price} process of the form (\ref{Eq:DfltSecDf}), where $\mathcal{X} = \varrho(S_T)$, for a deterministic function $\varrho$.
We denote it by $\Pi(t;s) = (\Pi_1(t;s), \ldots, \Pi_N(t;s)){\Blue '}$, where
\begin{eqnarray*}
\Pi_i(t;s) &:=&  \Ex^{\Qx} \left[\varrho(S_T) e^{-\int_{t}^{T}(r_{s} + h_{s}L_{s})ds} \bigg| X_t = e_i, S_t =s\right].
\label{eq:vulnPi}
\end{eqnarray*}
{\Blue The proof is presented in Appendix section.}

\begin{theorem}\label{Th:FKRNDynCall}
Assume that {\Blue each $\Pi_i(t;s)$} is {\Blue continuously} differentiable in {\DBlue $t\in(0,T)$} and twice {\Blue continuously} differentiable in {\DBlue $s\in(0,\infty)$}. Then, the following statements hold:
\begin{enumerate}
	\item[{\bf (1)}] The vector of pre-default prices $\Pi(t;s) = \left(\Pi_1(t;s),\ldots,\Pi_N(t;s) \right){\Blue '}$ {\DMagenta satisfies} the following coupled system of {\DMagenta backward Partial Differential Equations (PDE)}, {\DMagenta in the domain {\DBlue $ 0 \leq t <T$}, $0 < s < \infty$:}
\begin{eqnarray}
 \nonumber & & \frac{\pa \Pi_i(t;s)}{\pa t} + r_i \left(s \frac{\pa \Pi_i(t;s)}{\pa s} - \Pi_i(t;s)\right) -(h_i L_i - a_{i,i}^{\Qx}(t)) \Pi_i(t;s) +  \frac{1}{2} \sigma_i^2 {\YGreen s^2} \frac{\pa^2\Pi_i(t;s)}{\pa^2s} + \sum_{j \neq i} a_{i,j}^{\Qx}(t) \Pi_j(t;s) = 0 \\
 & & \gamma_i(T,s) = \varrho(s) , \qquad i = 1,\ldots,N.
\label{eq:callpdeGen}
\end{eqnarray}
\item[{\bf (2)}] The prices process (\ref{Eq:DfltSecDfNet}), denoted {\DBlue hereafter} by {$\Gamma(t,T;S_t)$}, of the vulnerable claim with payoff $\mathcal{X} = \varrho(S_{T})$ {\Awesome has the following {\DBlue semimartingale} representation}, for $ 0\leq{}t<T$, $ 0 < S_t < \infty$, {\Awesome $t < \tau$}:
\begin{align}\label{DRNM0call}
d \,\Gamma(t,T;S_t)&=\Gamma(t^{-},T;S_{t-})\bigg( [r_t + h_t (L_t-1)] dt + \frac{{\left<{\Pi(t;{\Awesome S_{t}})}, dM^{\Qx}(t)\right>}}{{\left<{\Awesome \Pi(t;S_{t}), X_{t^-}}\right>}}
+ {\Awesome \sigma_t} S_t {\Awesome \frac{\frac{\pa \left<\Pi(t;S_t), X_t \right>}{\pa s}}{\left<{\Awesome \Pi(t;S_{t}), X_{t^{^-}}}\right>}} dW^{\Qx}_t - d\xi_t^{\Qx} \bigg)
\end{align}
under the risk-neutral probability measure $\Qx${\DBlue .}
\item[{\bf (3)}] { The prices process (\ref{Eq:DfltSecDfNet}) of the vulnerable claim with payoff $\varrho(S_{T})$ {\Awesome has the following {\DBlue semimartingale} representation}, for $0\leq{}t<T$, $ 0 < S_t < \infty$, {\Awesome $t < \tau$}:
\begin{eqnarray}\label{DRWP0call}
\nonumber d\,\Gamma(t,T;S_t) &=& \Gamma(t-,T;S_{t-}) \bigg( \bigg[r_t + h_t (L_t-1)+{D(t,S_t)} + \frac{S_t \frac{\pa \left<\Pi(t;S_t),X_t\right>}{\pa s}}{\left<{ {\Awesome \Pi(t;S_{t})},X_{t^{-}}} \right>}  \left( \mu_t - r_t \right) \bigg] dt \\
        & &\qquad+\;\frac{\left<{{\Awesome \Pi(t;S_{t})}}, dM^{\Px}(t)\right>}{\left<\Pi(t;S_{t}), X_{t^{-}}\right>} + {\Awesome \sigma_t} S_t \frac{ \frac{\pa \left< \Pi(t;S_t),X_t \right>}{\pa s}} {\left<{\Awesome \Pi(t;S_{t})},X_{t^{-}} \right>} dW_t -d\xi^{\Px}_{t} \bigg ),
\end{eqnarray}
under the historical probability measure $\Px$, where $D(t,S_t):=\left< (D_{1}(t,S_t),\dots,D_{N}(t,S_t))',X_{t}\right>$ with}
\begin{equation}\label{NDfnDf}
	{ D_{i}(t,S_t):=\sum_{j=1}^{N} (a_{i,j}(t)-a_{i,j}^{\Qx}(t))\frac{\Pi_{j}(t;S_t)}{\Pi_{i}(t;S_t)}
	= \sum_{j\neq i} (a_{i,j}(t)-a_{i,j}^{\Qx}(t))\left(\frac{\Pi_{j}(t;S_t)}{\Pi_{i}(t;S_t)}-1\right).}
\end{equation}
\end{enumerate}
\end{theorem}
}

\section{{\Blue Novel algorithm for pricing vulnerable claims}}\label{sec:vulncl}
This section introduces a novel algorithm for pricing vulnerable {\Blue claims. {\DBlue Concretely,} Section} \ref{sec:BscMthd} {\DBlue presents} the basic algorithm and {\DRed analyzes its computational complexity. 
Section \ref{sec:erra} {\DBlue develops} an error analysis {\DBlue of the algorithm by providing} precise {\DBlue bounds for approximation error of the method}.
Section \ref{sec:DcmpPrc} {\DBlue then} builds on the basic algorithm and {\DBlue extends it to the pricing} self-decomposable claims on the underlying chain. Section \ref{Sec:PrcVulnClm} introduces a novel algorithm to price vulnerable European type options using the basic algorithm.}

\subsection{{\Blue Markov chain dependent claims}}\label{sec:BscMthd}
In this section, we develop a novel {efficient} algorithm for pricing {a claim whose payoff depends on the underlying economic regime in place $\{X_{t}\}_{t\geq{}0}$. This algorithm is used to price simple European {\Blue claims} of the form $\Xi(X_{T})$ for a deterministic function $\Xi$ as in Section \ref{Sec:FKSerRepr}. {\Blue Let us consider the pre-default price function (\ref{BndPrc3}) under the $i^{th}$ regime. Our approach to compute (\ref{BndPrc3}) builds on the series representation (\ref{CmpPsii}). Concretely, we will derive} a formula for $\Phi_{i,m}[\Xi](\zeta)$ when the risk-neutral generator $A^{\Qx}$ is time-invariant. The formula will be expressed in terms of the Laplace transform of the ``symmetric" Dirichlet distribution, {defined by}
{\Awesome
\begin{equation}\label{DrcLap}
	\mathcal{L}_{m}(x_{1},\dots,x_{m}) :=m! \int_{T_{m}}e^{-\sum_{j=1}^{m}x_{j}\lambda_{j}}d\lambda,
\end{equation}
where {$T_{m}:= \{(\lambda_{1},\dots,\lambda_{m}) \in \Rx^{m}: \lambda_{i}\geq 0, \; \sum_{i=1}^{m} \lambda_{i}\leq 1\}$}. 
}
The proof of the following result is given in {\DBlue Appendix} \ref{NmrMthdApdx}.
\begin{proposition}\label{MnLmItrFrm}
	Suppose that $a_{i,j}^{\Qx}(t)\equiv a_{i,j}^{\Qx}$  (hence, $\widetilde{K}$ and $\widetilde{r}$ are also time-invariant). Then, for $m\geq{}1$, we have that
	\begin{align}\label{CmpFrmPhi}
	{\Phi_{i,m}[\Xi](\zeta) = \frac{1}{(N-1)^{m}}\sum_{(\tilde{e}_{1}, \dots, \tilde{e}_{m})}  \Xi(\tilde{e}_{m})
e^{-\zeta {\tilde{r}'} \tilde{e}_{m} - \sum_{n=1}^{m}\tilde{e}_{n-1}^{'} \widetilde{K} \tilde{e}_{n}}
{\mathcal{L}_{m}(\zeta{\tilde{r}'}(\tilde{e}_{0} -\tilde{e}_{m}), \dots, \zeta{\tilde{r}'}(\tilde{e}_{m-1} - \tilde{e}_{m}))},}
\end{align}
where {\DBlue we set} {$\tilde{e}_{0}=e_{i}$ and} the {above} summation is over all ``paths" $(\tilde{e}_{1}, \dots, \tilde{e}_{m})$ such that $\tilde{e}_{j}\in\{e_{1},\dots,e_{N}\}$ and $\tilde{e}_{j}\neq \tilde{e}_{j-1}$, for $j=1,\dots,m$.
\end{proposition}

\begin{remark}\label{PartCas}
Note that
\begin{align}\label{ExctValPhi1Phi2}
	{\Phi_{i,0}[\Xi](\zeta)={\Awesome \Xi(e_i)} e^{-\zeta \tilde{r}_{i}}},\qquad \Phi_{i,1}[\Xi](\zeta)=\frac{1}{(N-1)}\sum_{j\in\{1,\dots,N\}\backslash\{i\}}{\Awesome \Xi(e_j)}
	e^{-\zeta \tilde{r}_{j}-\widetilde{K}_{i,j}}\frac{1}{\zeta(\tilde{r}_{i}-\tilde{r}_{j})}\left(1-e^{-\zeta(\tilde{r}_{i}-\tilde{r}_{j})}\right).
\end{align}
For a general $m$, the following Taylor approximation around the origin {will turn out to be quite} useful to compute the Dirichlet Laplace transform (\ref{DrcLap}):
\begin{equation}\label{TylrApr}
{\mathcal{L}_{m}(x_{1},\dots,x_{m}) = 1 - \frac{\sum_{i=1}^{m}x_{i}}{m+1} + \frac{\sum_{i=1}^{m}x_{i}^{2} + \frac{1}{2}\sum_{i \neq j}x_{i}x_{j}}{(m+1)(m+2)} + O\left(\|(x_{1},\dots,x_{m}) \|^{3/2}\right)}.
\end{equation}
\end{remark}

{In order to evaluate the option price}, the infinite series (\ref{CmpPsii}) will be truncated and the Laplace transform in (\ref{CmpFrmPhi}) may {also} be approximated by (\ref{TylrApr}). Both of these approximations are valid when time-to-maturity {$\zeta=T-t$} is small. Hence, {it useful to express the option price (\ref{OptPrc2}) in terms of the price of {options with {\DBlue shorter} expiration.}} Concretely, fix a small mesh {\Blue parameter} $\delta:=T/k$ for a {\Blue given} positive integer $k$ and let
\begin{equation}\label{DfIuv}
	I_{u,v}:={-\int_{u}^{v}{\tilde{r}'}X_{s}ds - \displaystyle{\sum_{s \in (u,v]: \Delta X_{s} \neq 0}}X_{s^{-}}'\tilde{K}X_{s}}.
\end{equation}
Then, using {the tower and Markov properties}, the time $0$ price of the vulnerable claim may be computed as follows
\begin{align} 
\nonumber \Psi_{i}[\Xi](0;T)&=\Ex^{\widetilde{\Qx}}\left[\left.e^{I_{0,T}} \Xi(X_{T})\right| X_{0}=e_{i}\right]= \Ex^{\widetilde{\Qx}}\left[\left.e^{I_{0,\delta}} \Ex^{\widetilde{\Qx}}\left[\Xi(X_{T}) e^{I_{\delta,T}}|\mathcal{F}_{\delta}\right]\right| X_{0} = e_{i} \right]  \\
 &=: \Ex^{\widetilde{\Qx}}\left[\left. e^{I_{0,\delta}}\widetilde{\Xi}(X_{\delta})\right| X_{0} = e_{i} \right] =\Psi_{i}[\widetilde{\Xi}](0;\delta), \label{SmallTimeExp}
\end{align}
where {\Blue we introduced a new} payoff function $\widetilde{\Xi}$ defined as $\widetilde{\Xi}(e_{j})=\Psi_{j}[\Xi](0;T-\delta)$. Note that {\Blue the last expression in} (\ref{SmallTimeExp}) {\Blue represents} the price of an option with {short} maturity $\delta$ and, hence, it can be accurately computed by taking $M$ terms in (\ref{CmpPsii}) and (possibly) using (\ref{TylrApr}). {In} order to evaluate the option's payoff {$\widetilde{\Xi}$}, we {apply} again the procedure (\ref{SmallTimeExp}), replacing $T$ by $T-\delta$.
Computationally, {\Blue one} can create a recursive or iterative implementation in order to evaluate (\ref{SmallTimeExp}). {The pseudo-code of the proposed iterative algorithm is given in Appendix \ref{Sec:PseudoCodesAll}} {(Algorithm \ref{alg:FindOptPrice} therein).}


\begin{remark}\label{Rem:TDGConj}
{\Blue
	The previous {method} can be adapted to deal with sufficiently smooth time dependent functions $\tilde{r}(t)$ and $\tilde{K}(t)$ (e.g., continuously differentiable in $[0,T]$). Indeed, as in (\ref{SmallTimeExp}) {and using {notation} (\ref{DfnCFunc})}, we have
	\begin{equation}
	\Psi_{i}[\Xi](t;T)=\Ex^{\widetilde{\Qx}}\left[\left. e^{I_{t,T}}\Xi(X_{T})\right| X_{t}=e_{i}\right]= \Ex^{\widetilde{\Qx}}\left[\left.e^{I_{t,t+\delta}}\Psi_{C_{t+\delta}}[\Xi](t+\delta;T)\right| X_{t} = e_{i} \right]=\Psi_{i}[\widetilde{\Xi}](t;t+\delta), \label{SmallTimeExp2}
\end{equation}
fixing $\widetilde{\Xi}(e_{i}):=\Psi_{i}[\Xi](t+\delta;T)$.
Then, for sufficiently small $\delta$, $\Psi_{i}[\Xi](t;T)$ can be accurately computed as if $\tilde{r}$ and $\tilde{K}$ were time-invariant during the period $[t,t+\delta]$. We analyze the error of this approach in the section \ref{SSec:ErrTDG} below (see also the end of Section \ref{SSec:ExactPhis} for another approach).}
\end{remark}

\begin{remark}
	The complexity of the proposed method (Algorithm \ref{alg:FindOptPrice} in Appendix \ref{Sec:PseudoCodesAll}) can be evaluated as follows. Let
\begin{equation}\label{CmpPsiiM}
	{\Blue \widetilde{\Psi}_{i}^{(M)}[\Xi](t;T):=\sum_{m=0}^{M-1} e^{-(T-t)}\frac{(T-t)^{m}}{m!}\Phi_{i,m}[\Xi](T-t)}.
\end{equation}
The computational complexity to evaluate {\Blue $\widetilde{\Psi}_{i}^{(M)}[\Xi](t;T)$ for each $i$} is O($N^{M-1}$). {Indeed}, for each $m$, it is required to evaluate $\Phi_{i,m}[\Xi](T-t)$. This, in turn, requires at most $N^{M-1}$ evaluations of the Laplace transform of the symmetric Dirichlet distribution {(one for each path {$(\tilde{e}_{0},\dots,\tilde{e}_{m})$} {\DRed such} that $\tilde{e}_{0}=e_{i}$ and $\tilde{e}_{j}\neq\tilde{e}_{j-1}$ for $j=1,\dots,m$  as seen in (\ref{CmpFrmPhi}))}. Therefore, the total computational complexity {\Blue to evaluate (\ref{CmpPsiiM})} is $O(N^{M-1})$ and, thus, the algorithm has complexity $O(N^{M})$ to compute the prices conditional on all starting regimes. We also remark that, for a fixed $M$, the computation of $\Psi_{i}[\Xi](0;T)$ can be {sped} up greatly by  saving the paths $(\tilde{e}_{1},\dots,\tilde{e}_{m})$ needed for (\ref{CmpFrmPhi}) at the beginning and reusing them for each evaluation of $\Phi_{i,m}[\Xi](\zeta)$ with $\zeta\in\{\delta, \dots, k\delta\}$.
\end{remark}

{\Blue
\subsection{Error analysis} \label{sec:erra}
We now proceed to give precise error bounds for our proposed method. First, we need to extend the framework of our algorithm.
	In this section we fix positive integers $M$ and $k$, and an integer $r\in\{1,\dots,M\}$. We also let $\delta=T/k$ and ${\bf 1}(e_{i})=1$, for all $i=1,\dots,N$. Suppose the following conditions hold true:
	 \begin{enumerate}
	 	\item[{\bf (1)}] The functionals
	 \[
	 	\Phi_{i,0}[\Xi](u;v),\dots, \Phi_{i,r-1}[\Xi](u;v)
	\]
	 defined in (\ref{AuxDfn1b}) can be computed exactly for any payoff function $\Xi:\{e_{1},\dots,e_{N}\}\to\Rx$ and any $0\leq{}u\leq{}v\leq{}T$.
	 	\item[{\bf (2)}] For any $m\in \{r,\dots,M-1\}$, $i\in\{1,\dots,N\}$, $\Xi:\{e_{1},\dots,e_{N}\}\to\Rx$, and $0\leq{}\delta\leq{}T$,  there is an approximation  $\widetilde{\Phi}_{i,m}[\Xi](\delta;T)$ of ${\Phi}_{i,m}[\Xi](\delta;T)$, linear in $\Xi$, such that 		
		\begin{align}\label{Eq:ApprxPhiCond}
			&{\Blue \left|\widetilde{\Phi}_{i,m}[\Xi](\delta;T)-{\Phi}_{i,m}[\Xi](\delta;T)\right|\leq{}B  \max_{\ell}|\Xi(e_{\ell})| \,\delta^{j}}, \quad \text{for all }\delta<\delta_{0},\\
			&\left|\widetilde{\Phi}_{i,m}[\Xi](\delta;T)\right|\leq{}\Phi_{i,m}[{\bf 1}](\delta;T)\max_{\ell}|\Xi(e_{\ell})|,\quad  \text{for all }\delta<\delta_{0},
			\label{Eq:ApprxPhiCond2}
		\end{align}
	for some $0<\delta_{0}\leq T$ and  $j\geq{}1$ and a constant $B<\infty$, independent of $\delta$, $i$, $m$, and $\Xi$.
\end{enumerate}
Next, define the following approximation for the pre-default payoff function ${\Psi}_{i}[\Xi](t;T)$:
	 \begin{equation}\label{Eq:N2layerApr}
	 	\widetilde{\Psi}_{i}^{(M,r)}[\Xi](t;T)=\sum_{m=0}^{r-1} e^{-(T-t)}\frac{(T-t)^{m}}{m!}\Phi_{i,m}[\Xi](T-t;T)+
		\sum_{m=r}^{M-1} e^{-(T-t)}\frac{(T-t)^{m}}{m!}\widetilde{\Phi}_{i,m}[\Xi](T-t;T).
\end{equation}
Note that the case $r=M$ essentially amounts to approximating the infinite series approximation (\ref{CmpPsii}) by the summation of the {\DRed first $M$ terms} and, hence, that the $M$ terms $\Phi_{i,0}[\Xi](u;v),\dots, \Phi_{i,M-1}[\Xi](u;v)$ are computed exactly.

	 Consider the following generalized algorithm to approximate $\Psi[\Xi](0;T):=(\Psi_{1}[\Xi](0;T),\dots,\Psi_{N}[\Xi](0;T))'$, which extends our approach in Section \ref{sec:BscMthd}:
	 \begin{enumerate}
	 	\item[{\bf (A)}]  {\DBlue Approximate} {\DViolet $\Psi_{i}[{\DBlue {\Xi}}]((k-1)\delta;T)$} by the functional $\widetilde{\Psi}_{i}^{(M,r)}[{\DBlue {\Xi}}]((k-1)\delta;T)$ in (\ref{Eq:N2layerApr}); denote this first step approximation $\widetilde{\Psi}^{(M,r,k)}[{\DBlue {\Xi}}]((k-1)\delta;T)$;
		\item[{\bf (B)}] For each $j$ from $k-2$ to $0$ (going backwards), compute iteratively
		\begin{equation}\label{Eq:BkwrdDfnPsiApp}
			 \widetilde{\Psi}^{(M,r,k)}[{\Xi}](j\delta;T)=\widetilde{\Psi}^{(M,r)}[\widetilde{\Xi}^{(j)}](j\delta;(j+1)\delta),
		\end{equation}	
		where ${\DBlue \widetilde{\Xi}^{(j)}}$ is computed as
		\[
			\widetilde{\Xi}^{(j)}(e_{i}):= \widetilde{\Psi}^{(M,r,k)}_{i}[{\Xi}]((j+1)\delta;T).
		\]
	 \end{enumerate}
{\Blue	
The following result will be important in the subsequent error analysis. Its proof is given in the Appendix \ref{NmrMthdApdx}}.
\begin{theorem}\label{Lem:MnEstErr}
	 Under the above conditions {\bf (1)-(2)}, the previous algorithm {\bf (A)-(B)} will result in a price approximation $\widetilde{\Psi}_{i}^{(M,r,k)}[\Xi](0;T)$ such that
	\begin{equation}\label{Eq:KIFEAn}
		 \max_{i\in\{1,\dots,N\}}\left|{{\Psi}_{i}[\Xi](0;T)-{\DViolet \widetilde{\Psi}_{i}^{(M,r,k)}}[\Xi](0;T)}\right|\leq{} D \frac{T^{\alpha}}{k^{\alpha-1}},
	\end{equation}
	where {\Blue $\alpha:=(r+j)\wedge M$ and $D=\|\Xi\|_{\infty}(1+B)$} with $B$ defined as in (\ref{Eq:ApprxPhiCond}) and  $\|\Xi\|_{\infty}:=\max_{\ell=1,\dots,N}|\Xi(e_{\ell})|$.
\end{theorem}

The following result relaxes the condition (\ref{Eq:ApprxPhiCond2}), but does not yield {\Blue a bound as explicit} as in (\ref{Eq:KIFEAn}).
\begin{corollary}\label{Lem:MnEstErrMod}
	 Suppose that the above conditions {\bf (1)-(2)} hold true, but replacing (\ref{Eq:ApprxPhiCond2}) with the following milder condition:
	 \begin{align}\label{Eq:ApprxPhiCond2Milder}
	 	\left|\widetilde{\Phi}_{i,m}[\Xi](\delta;T)\right|\leq{}C\max_{\ell}|\Xi(e_{\ell})|,\quad  \text{for all }\delta<\delta_{0},
	 \end{align}
	 for some $0<\delta_{0}\leq T$ and a constant $C<\infty$, independent of $\delta$, $i$, $m$, and $\Xi$. Then, there exists a constant $E$ such that, for all $k\geq{}1$,
	\begin{equation}\label{Eq:KIFEAnMod}
		 \max_{i\in\{1,\dots,N\}}\left|{{\Psi}_{i}[\Xi](0;T)-\widetilde{\Psi}_{i}^{(M,r,k))}[\Xi](0;T)}\right|\leq{} E \frac{T^{\alpha}}{k^{\alpha-1}},
	\end{equation}
	where $\alpha:=(r+j)\wedge M$.
\end{corollary}
We now consider different applications of our previous general approximation framework.
\subsubsection{Error analysis when {\Blue the} $\Phi_{i,m}$'s are exact}\label{SSec:ExactPhis}
As a direct consequence of {\DRed Theorem \ref{Lem:MnEstErr}}, by taking {\Blue $r=M$ and $B=0$}, the approximation
\[
	\widetilde{\Psi}_{i}^{(M)}[\Xi](t;T):=\sum_{m=0}^{M-1} e^{-(T-t)}\frac{(T-t)^{m}}{m!}\Phi_{i,m}[\Xi](T-t),
\]
attains the following error:
	\[
		 \max_{i\in\{1,\dots,N\}}\left|{{\Psi}_{i}[\Xi](0;T)-\widetilde{\Psi}_{i}^{(M)}[\Xi](t;T)}\right|\leq{} \|\Xi\|_{\infty}\times \frac{T^{M}}{k^{M-1}}.
	\]
	As indicated in the Remark \ref{PartCas}, for time-invariant generators, {\Blue $\Phi_{i,0}[\Xi]$ and $\Phi_{i,1}[\Xi]$} can be computed {\Blue in closed form}. Hence, using only these two values, the bond prices can be {\Blue evaluated} up to an error of order $T^{2}/k$ using $k$ iterations {\Blue and} with a maximal polynomial complexity of $O(N^{3})$. {\Blue Furthermore, even in the case of time-dependent generators, the functionals $\Phi_{i,0}[\Xi](\zeta;T)$ and $\Phi_{i,1}[\Xi](\zeta;T)$  can be computed by the expressions (\ref{Eq:1st2ndCoeff}) (at the cost of computing a single integral) and, thus, the same error will hold true. This last observation {\Blue partly} justifies our claim in {\Blue Remark \ref{Rem:TDGConj}}.}
	
\subsubsection{Error analysis for smooth time-dependent generators}\label{SSec:ErrTDG}
In the case of time-dependent generators, {\DBlue a natural approach is to} approximate ${\Phi}_{i,m}[\Xi](\delta;T)$ when $\delta$ is small by freezing the values of $\tilde{r}$ and $\tilde{K}$ at the beginning of the integration interval $[T-\delta,T]$. Concretely, consider the approximation
\begin{equation}\label{Eq:PrpApprTD}
	\widetilde{\Phi}_{i,m}[\Xi](\delta;T):=\Ex^{\widetilde{\Qx}}_{i}\left[ \Xi(\widetilde{X}_{m})\exp\left\{
	-\sum_{n=0}^{m} \int_{\delta U_{(n)}}^{\delta U_{(n+1)}}{\tilde{r}(T-\delta)'} \widetilde{X}_{n} ds-\sum_{n=1}^{m}\widetilde{X}_{n-1}'\widetilde{K} (T-\delta) \widetilde{X}_{n}\right\}\right].
\end{equation}
These approximations can be expressed in closed {\YGreen form} in terms {\Blue of} ``symmetric" Dirichlet {\Blue Laplace} transforms as in Proposition \ref{MnLmItrFrm}. Obviously,  (\ref{Eq:PrpApprTD}) satisfies {\Blue (\ref{Eq:ApprxPhiCond2Milder})}. In order to verify the validity of (\ref{Eq:ApprxPhiCond}), let us {\DRed assume} that for all $p,q$, the functions $\tilde{r}_{p}$ and $\widetilde{K}_{p,q}$ are continuously differentiable on $[0,T]$. Denoting the random variable inside the expectation on the right-hand side of (\ref{Eq:PrpApprTD}) by $G$ and from the definitions {\Blue (\ref{AuxDfn1b})} and (\ref{Eq:PrpApprTD}),  note that
\begin{align*}
	&\left|\widetilde{\Phi}_{i,m}[\Xi](\delta;T)-{\Phi}_{i,m}[\Xi](\delta;T)\right|\\
	&\quad\leq{}\Ex^{\widetilde{\Qx}}_{i}\bigg[ G\bigg|1-e^{
	-\sum_{n=0}^{m} \int_{\delta U_{(n)}}^{\delta U_{(n+1)}}[{\tilde{r}(T-\delta+s)'}-{\tilde{r}(T-\delta)'}] \widetilde{X}_{n} ds -\sum_{n=1}^{m}\widetilde{X}_{n-1}'[\widetilde{K} (T-\delta+\delta U_{(n)})-\widetilde{K} (T-\delta)] \widetilde{X}_{n}}\bigg|\bigg]\\
	&\quad\leq{}\Ex^{\widetilde{\Qx}}_{i}\bigg[ G\bigg|1-e^{
	-\delta^{2}\max_{p}\sup_{{\Blue s\in [T-\delta,T]}}|\tilde{r}'_{p}(s)| - m \delta
	\max_{p\neq q}\sup_{{\Blue s\in [T-\delta,T]}}|\widetilde{K}'_{p,q}(s)| }\bigg|\bigg]\\
	&\leq B \|\Xi\|_{\infty}\delta,
\end{align*}
for some constant $B$ independent of $\Xi$. Thus, (\ref{Eq:ApprxPhiCond}) is satisfied with $j=1$. In particular, if we use the approximation (\ref{Eq:PrpApprTD}) for $m=1,\dots,M-1$ (so that $r=1$), then {\DBlue the parameter $\alpha$ of Theorem \ref{Lem:MnEstErr} will be two} and the resulting pre-default claim price approximation will attain an error bound of the form:
	\[
		 \max_{i\in\{1,\dots,N\}}\left|{{\Psi}_{i}[\Xi](0;T)-\widetilde{\Psi}_{i}^{(M)}[\Xi](t;T)}\right|\leq{} E  \frac{T^{2}}{k}.
	\]

\subsubsection{Error analysis arising from {\DBlue the} Taylor approximations}	

Another natural issue is to {\DBlue quantify and control} the error arising from using a Taylor approximation to the Dirichlet Laplace transforms (\ref{DrcLap}) as suggested in Remark \ref{PartCas}. Concretely, let us recall the following expression from {Proposition} \ref{MnLmItrFrm},
	\begin{align}\label{CmpFrmPhi22}
	{\Phi_{i,m}[\Xi](\zeta) = \frac{1}{(N-1)^{m}}\sum_{{\overset{\left( \tilde{e}_{1},\dots,\tilde{e}_{m} \right)}{\tilde{e}_{i} \neq \tilde{e}_{i+1}} }}   \Xi(\tilde{e}_{m})
e^{-\zeta {\tilde{r}'} \tilde{e}_{m} - \sum_{n=1}^{m}\tilde{e}_{n-1}^{'} \widetilde{K} \tilde{e}_{n}}
{\mathcal{L}_{m}(\zeta{\tilde{r}'}(\tilde{e}_{0} -\tilde{e}_{m}), \dots, \zeta{\tilde{r}'}(\tilde{e}_{m-1} - \tilde{e}_{m}))},}
\end{align}
valid for time-invariant generators. Next, we consider the following ``Taylor approximation":
	\begin{align}\label{CmpFrmPhiTaylor}
	{\widetilde{\Phi}_{i,m}[\Xi](\zeta) = \frac{1}{(N-1)^{m}}\sum_{{\overset{\left( \tilde{e}_{1},\dots,\tilde{e}_{m} \right)}{\tilde{e}_{i} \neq \tilde{e}_{i+1}} }}  \Xi(\tilde{e}_{m})
e^{-\zeta {\tilde{r}'} \tilde{e}_{m} - \sum_{n=1}^{m}\tilde{e}_{n-1}^{'} \widetilde{K} \tilde{e}_{n}}
{\widehat{\mathcal{L}}_{m}(\zeta{\tilde{r}'}(\tilde{e}_{0} -\tilde{e}_{m}), \dots, \zeta{\tilde{r}'}(\tilde{e}_{m-1} - \tilde{e}_{m}))},}
\end{align}
where we approximate the Laplace transform (\ref{DrcLap}) by the following Taylor polynomial of $p^{th}$-order:
{\Awesome
\begin{equation*} \label{TaylorLaplace}
\widehat{\mathcal{L}}_{m}(x_{1},\dots,x_{m}):=\widehat{\mathcal{L}}^{(p)}_{m}(x_{1},\dots,x_{m}):= m! \sum_{\ell=0}^{p} \int_{T_{m}} \frac{\left< \mathbf{x},\mathbf{\lambda} \right> ^{\ell}}{\ell!} d\lambda.
\end{equation*}
}
Let us now verify the {\DBlue conditions} for {\DBlue Corollary} \ref{Lem:MnEstErrMod}. It is easy to {\DViolet verify} that {$\widetilde{\Phi}_{i,m}[\Xi]$} satisfies the condition {\Blue (\ref{Eq:ApprxPhiCond2Milder})} 
. It remains to show the validity of (\ref{Eq:ApprxPhiCond}). First note that, by the Cauchy-Schwarz's and Jensen's inequalities,
\begin{align*}
|\mathcal{L}_{m}(x_{1},\dots,x_{m}) - \widehat{\mathcal{L}}^{(p)}_{m}(x_{1},\dots,x_{m})| &\leq m! \sum_{i=p+1}^{\infty} \int_{T_{m}} \frac{\left| \left< \mathbf{x},\mathbf{\lambda} \right> \right|^{i}}{i!} d\lambda \leq  m! \sum_{i=p+1}^{\infty}  \frac{\| x \|^{i}}{i!} \int_{T_{m}} \| \lambda \|^{i}  d\lambda\\
&\leq  m! \sum_{i=p+1}^{\infty}  m^{i/2-1}\frac{\| x \|^{i}}{i!} \int_{T_{m}} \left(\sum_{\ell=1}^{m}\lambda_{\ell}^{i}\right)  d\lambda\\
&\leq   \sum_{i=p+1}^{\infty}  m^{i/2}\frac{\| x \|^{i}}{i!}\frac{m!\,i!}{(m+i)!},
\end{align*}
where we had also used the fact that  {\Awesome $m! \int_{T_{m}}\lambda_{\ell}^{i} d\lambda$} coincides with $i^{th}$ moment of a ${\rm Beta}(1,m)$ random variable, which is known to be $m!i!/(m+i)!$. Using the previous estimate, it follows that
\begin{equation}\label{Eq:NdBndTAMT}
|\widetilde{\Phi}_{i,m}[\Xi](\zeta) - \Phi_{i,m}[\Xi](\zeta)| \leq K \frac{\zeta^{p+1}}{(N-1)^{m}}
\displaystyle{\sum_{ \overset{\left( \tilde{e}_{1},\dots,\tilde{e}_{m} \right)}{{\tilde{e}_{i} \neq \tilde{e}_{i+1}}}}} \Xi(\tilde{e}_{m})e^{-\zeta \tilde{r}'\tilde{e}_{m} {-} \sum_{n=1}^{m} \tilde{e}_{n-1}^{'} \tilde{K}\tilde{e}_{n} } =: K \zeta^{p+1} \widetilde{C}_{m},
\end{equation}
for an absolute constant $K$.  Finally, its is clear that
\begin{align*}
\widetilde{C}_{m} & = \frac{1}{(N-1)^{m}}
\displaystyle{\sum_{ \overset{\left( \tilde{e}_{1},\dots,\tilde{e}_{m} \right)}{{\tilde{e}_{i} \neq \tilde{e}_{i+1}}}}} \Xi(\tilde{e}_{m})e^{-\zeta \tilde{r}\tilde{e}_{m} {-} \sum_{n=1}^{m} \tilde{e}_{n-1}^{'} \tilde{K}\tilde{e}_{n}},
\end{align*}
can be bounded by $\|\Xi\|_{\infty} R$, for some constant $R$. In light {\DRed of} (\ref{Eq:NdBndTAMT}), {\DRed Corollary} \ref{Lem:MnEstErrMod} implies that the previous Taylor-based approximations will attain an error of the form
	\begin{equation}
		 \max_{i\in\{1,\dots,N\}}\left|{{\Psi}_{i}[\Xi](0;T)-\widetilde{\Psi}_{i}^{(M,r,k))}[\Xi](0;T)}\right|\leq{} E \frac{T^{\alpha}}{k^{\alpha-1}},
	\end{equation}
with {$\alpha:=r+ (p+1) = (3+p)\wedge M$} if we take $M>{}2$ and use the exact values (\ref{ExctValPhi1Phi2}) for $m\in\{0,1\}$ and Taylor approximations of $p^{th}$ order for $m=2,\dots,M$ (so that $r=2$). 	
}

\subsection{Pricing of decomposable claims on the underlying Markov process}\label{sec:DcmpPrc}
 We proceed to {\DBlue extend our approach above to price a class of} path-dependent claims, termed {self-decomposable claims}, whose payoffs can be decomposed into {shorter maturity payoffs}. We illustrate {\DBlue our} method for a type of barrier option on the Markov process $\{X_{t}\}_{t\geq{}0}$.
As it is evident from (\ref{SmallTimeExp}), our approach heavily relies on being able to decompose the payoff of the claim into payoffs of shorter maturity. The following broad definition attempts to give a more precise meaning to this concept:
\begin{definition}
Consider a family of payoffs $\{\Sigma_{t,T}\}_{0\leq{}t\leq{}T}$, where for each $0\leq{}t\leq{}T$, $\Sigma_{t,T}$ represents a payoff depending on the path of $X$ on $[t,T]$. We write $\Sigma_{t,T} := \Sigma(\{X_{s}\}_{t \leq s \leq T})$. We say that the family $\{\Sigma_{t,T}\}_{0\leq{}t\leq{}T}$ is {self-decomposable} if, for any {$0<t<t'<T$}, the following decomposition holds true:
\[
	\Sigma_{t,T} = {f(\Sigma_{t,{t'}})+ g(\Sigma_{t,{t'}})\Sigma_{t',T}},
\]
for some measurable functions {$f,g: \mathbb{R} \to \mathbb{R}$}.
\end{definition}

We now proceed to describe our {\DBlue method}. Following our strategy for simple claims, it is natural that a feasible {\DBlue procedure} to price a {self-decomposable} claim  $\Sigma_{t,T}(X) \equiv \Sigma(\{X_{s}\}_{t \leq s \leq T})$ will {consist} of {the} following two general steps:
\begin{description}
	\item[(Decomposition)] Fix $\delta=(T-t)/k$ for a positive integer $k$ and {apply} the following decomposition with $t':= t+\delta$:
		\begin{align*}
			\Ex^{\widetilde{\Qx}}\left[\left.e^{I_{t,T}} \Sigma_{t,T}\right| X_{t}=e_{i}\right]
			&= {\Ex^{\widetilde{\Qx}}\left[\left.e^{I_{t,t'}} f(\Sigma_{t,t'})\Ex^{\widetilde{\Qx}}[e^{I_{t',T}} |\mathcal{F}_{t'}]\right| {\Blue X_{t}} = e_{i} \right]+
			\Ex^{\widetilde{\Qx}}\left[\left.e^{I_{t,t'}} g(\Sigma_{t,t'})\Ex^{\widetilde{\Qx}}\left[\left.e^{I_{t',T}} \Sigma_{t',T} \right|\mathcal{F}_{t'}\right]\right| {\Blue X_{t}} = e_{i} \right]}\\
&=: {\Ex^{\widetilde{\Qx}}\left[\left.e^{I_{t,t'}} f(\Sigma_{t,t'})\Xi(X_{t'})\right| {\Blue X_{t}} = e_{i} \right]+
			\Ex^{\widetilde{\Qx}}\left[\left.e^{I_{t,t'}} g(\Sigma_{t,t'})\widetilde{\Xi}(X_{t'})\right| {\Blue X_{t}} = e_{i} \right]},
\end{align*}
{where $\Xi(e_{l}):=\Ex^{\widetilde{\Qx}}[e^{I_{t',T}} |X_{t'}=e_{l}]$ and ${\widetilde{\Xi}(e_{l})}:=\Ex^{\widetilde{\Qx}}\left[\left.e^{I_{t',T}} \Sigma_{t',T} \right|X_{t'}=e_{l}\right]$.  We then repeat the above decomposition to evaluate the payoffs {${\Xi}(\cdot)$} until $T-t'$ is small enough}.

	\item[(Near-expiration approximation)] {We proceed to apply an efficient approximation to evaluate {claims} of the form {$\Ex^{\widetilde{\Qx}}\left[\left.e^{I_{t,t'}} h(\Sigma_{t,t'}){\Xi}(X_{t'})\right| X_{t} = e_{i} \right]$}, when $t$ is close to $t'$.}
\end{description}

\begin{remark}
	The above method can also be extended to deal with claims whose payoffs can be decomposed in terms of the payoffs of other types of claims. For instance, we can say that two families of payoffs, say $\{\Sigma_{t,T}^{(0)}\}_{0\leq{}t\leq{}T}$ and $\{{\Sigma}^{(1)}_{t,T}\}_{0\leq{}t\leq{}T}$, are mutually self-decomposable if, for any $t<t'<T$,
\[
	\Sigma_{t,T}^{(k)} = f_{k}(\Sigma^{(0)}_{t,{t'}})+ g_{k}(\Sigma^{(0)}_{t,{t'}})\Sigma^{(0)}_{t',T}
	+h_{k}(\Sigma^{(1)}_{t,{t'}})+\ell_{k}(\Sigma^{(1)}_{t,{t'}})\Sigma^{(1)}_{t',T},
\]	
for each $k=0,1$, and some measurable functions $f_{k}, g_{k},h_{k},\ell_{k}$.
\end{remark}

As an illustration, we now consider the risk-neutral pricing of European barrier and digital contracts written on {the} volatility process $(\sigma_{t})_{t}$.  One may view the process $\sigma_{t}:=\sigma' X_{s}$ as a proxy to the volatility of a market index.  Instruments written on this process may be used to hedge volatility risk associated with periods of macro-economic bust or boom akin to that experienced by the U.S. economy leading into the 2008 crisis.
Let us define the following family of path-dependent payoffs:
\[
	\Sigma_{t,T} := {\bf 1}_{\big\{ \displaystyle{\max_{t \leq s \leq T}} \sigma' X_{s} \geq B\big\}}, \qquad \text{and}\qquad
	\widetilde{\Sigma}_{t,T} := {\bf 1}_{\big\{ \displaystyle{\max_{t \leq s \leq T}} \sigma' X_{s} < B\big\}}.
\]
The following simple relationships show that the family of payoffs $\{\Sigma_{t,T}\}_{0\leq{}t\leq{}T}$ {and $\{\widetilde{\Sigma}_{t,T}\}_{0\leq{}t\leq{}T}$ are} {self-decomposable}:
\begin{equation}\label{Eq:DcmpPrBr}
	{\rm (1)} \;\; \widetilde{\Sigma}_{t,T} = \widetilde{\Sigma}_{t,{t'}}\widetilde{\Sigma}_{{t'},T},
	\qquad {\rm (2)}\;\;{\Sigma_{t,T} \; = \; \Sigma_{t,{t'}} + {\Sigma_{t',T}}  - \Sigma_{t,{t'}}{\Sigma_{t',T}} = \Sigma_{t,{t'}}+(1 - \Sigma_{t,{t'}}){\Sigma_{t',T}}, \qquad (t<{}t'<T)}. 
\end{equation}
Now, let us consider the following {European} knock-out style barrier option:
\[
	\Psi_{i}^{OUT}[\Xi](t;T):=\Ex^{{\Qx}}\left[\left. e^{-\int_{t}^{T} r_{s} ds}
	\widetilde{\Sigma}_{t,T}\Xi(X_{T})\right| X_{t} = e_{i} \right]=
	\Ex^{{\widetilde{\Qx}}}\left[\left. e^{\hat{I}_{t,T}}
	\widetilde{\Sigma}_{t,T}\Xi(X_{T})\right| X_{t} = e_{i} \right],
\]
where we had used  the same change of probability  measure $\widetilde{\Qx}$ as in (\ref{OptPrc2}) and the following process analog to (\ref{DfIuv}):
\[
	\widehat{I}_{u,v}:= -\int_{u}^{v}{r'X_{s}}ds - {\sum_{s \in (u,v]: \Delta X_{s} \neq 0}} X_{s^{-}}'\tilde{K}X_{s}.
\]
Then, the following decomposition follows {from (\ref{Eq:DcmpPrBr})}:
\begin{align}\nonumber
	\Psi_{i}^{OUT}[\Xi](t;T)
	&= \Ex^{\widetilde{\Qx}}\left[e^{\widehat{I}_{t,{t'}}}\widetilde{\Sigma}_{t,{t'}} \left.\Ex^{\widetilde{\Qx}}\left[\left.\widetilde{\Sigma}_{{t'},T}\Xi(X_{T}) e^{\widehat{I}_{{t'},T}} \right| \mathcal{F}_{{t'}} \right] \right| X_{t} = e_{i} \right]\\
	&= \Ex^{\widetilde{\Qx}}\left[\left.e^{\widehat{I}_{t,{t'}}}\widetilde{\Sigma}_{t,{t'}} \Psi_{C_{t'}}^{OUT}[\Xi]({t'};T)\right| X_{t} = e_{i} \right]   = \Psi_{i}^{OUT}[\Psi^{OUT}_{C_{t'}}[\Xi]({t'};T)](t;{t'}).
\label{KO_Parity}
\end{align}
For a near-expiration approximation method for $\Psi_{i}^{OUT}[\Xi](t;t')$ (i.e. when $\zeta:=t'-t\approx 0$), we use again (\ref{CmpPsii}), {(\ref{AuxDfn1b}), and (\ref{CmpFrmPhi})}:
\[
	\Psi_{i}^{OUT}[\Xi](t;t')= e^{-\zeta}\sum_{m=0}^{M-1}\frac{\zeta^{m}}{m!} \Phi^{OUT}_{i,m}(\zeta),
\]
with
	\begin{align*}
	\Phi_{i,m}^{OUT}(\zeta) = \frac{1}{(N-1)^{m}}\sum_{{\overset{\left( \tilde{e}_{1},\dots,\tilde{e}_{m} \right)}{\tilde{e}_{i} \neq \tilde{e}_{i+1}} }}  \Xi(\tilde{e}_{m}){\bf 1}_{\{\max_{j}(\sigma'\tilde{e}_{j})<B\}}
e^{-\zeta {\tilde{r}'} \tilde{e}_{m} - \sum_{n=1}^{m}\tilde{e}_{n-1}^{'} \widetilde{K} \tilde{e}_{n}} \mathcal{L}_{m}(\zeta {\tilde{r}'}(\tilde{e}_{0} -\tilde{e}_{m}),  \dots, \zeta{\tilde{r}'}(\tilde{e}_{m-1} - \tilde{e}_{m})).
\end{align*}

Note that the knock-in style barrier option $\Psi_{i}^{IN}[\Xi](t;T):=\Ex^{{\Qx}} \left[ \exp \left \{-\int_{t}^{T} r_{s} ds \right \}
	{\Sigma}_{t,T}\Xi(X_{T})| X_{t} = e_{i} \right]$ can  be easily computed {using the relation} $\widetilde{\Sigma}_{t,T}=1-\Sigma_{t,T}$. {In} general one only needs to price either a knock-in or {a} knock-out contract{,} as the value of the other follows immediately from the knock-in/knock-out parity.
	

\subsection{Pricing of vulnerable call/put options}\label{Sec:PrcVulnClm}
{\DBlue We now develop} a new method to price vulnerable call/put options on the stock. The latter type of options have received growing interest in the literature {\Blue (see the introduction for further references).} {\DMagenta We set the terminal payoff $\varrho(S_T) = (S_T-K)^+$ in Eq.~\eqref{eq:vulnPi}, thus obtaining a vulnerable time $t$ call option price:
\[
	\Pi_i(t;s) := \Pi_i(t,T;s)=\Ex^{\Qx}\left[\left. (S_{T}-K)_{+}e^{-\int_{t}^{T}(r_{s} + h_{s}L_{s})ds} \right| X_{t} = e_{i},S_{t}=s \right].
\]
}
As before, we will change the probability measure into $\widetilde{\Qx}$ so that, in terms of the process $I_{t,T}$ defined in (\ref{DfIuv}),
\[
	{\DMagenta
	\Pi_i(t,T;s)=\Ex^{\widetilde{\Qx}}\left[\left. (S_{T}-K)_{+}e^{I_{t,T}} \right| X_{t} = e_{i},S_{t}=s \right].}
\]
{\DMagenta Under the risk-neutral measure defined by {\Blue Eqs.}~\eqref{eq:radonnik} and \eqref{eq:Qbrown}, we have that the {\DBlue stock price} is given by}
\[	
	{\DMagenta S_{T}=S_{t}\exp\left\{\int_{t}^{T} \tilde{b}' X_{s} ds + \int_{t}^{T} \sigma' X_{s} d {\DMagenta W^{\Qx}_{s}} \right\},}
\]
where the evolution of $X_t$ is determined by the risk-neutral generator $A^{\tilde{\Qx}}$ under $\tilde{\Qx}$ (see, e.g.,  \cite{elliottet}). Above,
 $\tilde{b}:=(\tilde{b}_{1},\dots,\tilde{b}_{N})'$ is given by $\tilde{b}_{i}:=r_{i}-\sigma_{i}^{2}/2$. In particular,  given $\sigma(X_{u}:t\leq{}u\leq{}T)$ and $S_{t}=s$, we can see $\{S_{u}\}_{t\leq{}u\leq{}T}$ as a geometric Brownian motion with a deterministic time-varying volatility and initial value $s$. As it is well-known, one can express the call price for such a model in terms of the Black-Scholes formula with constant volatility $\bar{\sigma}$, short-rate $r$,  spot price $s_0$, maturity $\zeta$, and strike $K$:
\[
	{\rm BS}\left(\zeta;s_0,\bar{\sigma}^{2},r,K\right):=e^{-r\zeta }\Ex^{\widetilde{\Qx}} \left(s_0 e^{\bar{\sigma} {\DMagenta W^{\Qx}_{\zeta}}+(r-\bar{\sigma}^{2}/2)\zeta} -K\right)_{+}.
\]
Concretely, denoting
\[
	\zeta:= T-t, \quad \bar{\sigma}^{2}:=\frac{1}{T-t}\int_{t}^{T}(\sigma' X_{u})^{2} du,\quad
	\bar{b}:=\frac{1}{T-t}\int_{t}^{T}\tilde{b}' X_{u} du,\quad
	s_0:= s e^{\bar{b} \zeta+\zeta\bar\sigma^{2}/2 }=s e^{\int_{t}^{T}r' X_{u} du},
\]
we have
\begin{align*}
	 \Ex^{\widetilde{\Qx}}\left[\left.(S_{T}-K)_{+}\right|\sigma(X_{u}:t\leq{}u\leq{}T),S_{t}=s\right] =
	 \Ex^{\widetilde{\Qx}}\left[(s e^{\bar{b} \zeta+\zeta\bar\sigma^{2}/2 }\times e^{\bar\sigma {\DMagenta W^{\Qx}_{\zeta}}-\zeta\bar\sigma^{2}/2}-K)_{+}\right]={\rm BS}(\zeta;s_{0},\bar{\sigma}^{2},0,K).
\end{align*}
Then, we obtain
\begin{align*}
	{\Pi_i(t,T;s)}&=\Ex^{\widetilde{\Qx}}\left[\left. \Ex^{\widetilde{\Qx}}\left[e^{I_{t,T}} \left.(S_{T}-K)_{+}\right|\sigma(X_{u}:t\leq{}u\leq{}T)\right] \right| X_{t} = e_{i},S_{t}=s \right]\\
	&=\Ex^{\widetilde{\Qx}}\left[\left. e^{I_{t,T}} {\rm BS}(\zeta;s_{0},\bar\sigma^{2},0,K)\right|X_{t} = e_{i},S_{t}=s \right].
\end{align*}
Let us now focus on the time-invariant case, where $A^{\tilde{\Qx}}$ is time-invariant and, hence, {\DMagenta $\Pi_i(t,T;s)= \Pi_i(0,T-t;s)$}. We set
\begin{align*}
	F(\zeta;s,e_{i})&:= \Ex^{\Qx}\left[\left.(S_{\zeta}-K)_{+}\right|S_{0}=s,X_{0}=e_{i}\right]=\Ex^{\widetilde{\Qx}}\Big[\left. e^{I_{0,\zeta}} {\rm BS}\Big(\zeta;se^{\int_{0}^{\zeta}r' X_{u}du},\zeta^{-1}\int_{0}^{\zeta}(\sigma' X_{u})^{2}du,0,K\Big)\right|S_{0}=s,X_{0} = e_{i} \Big].
\end{align*}
Our approach is based on two ideas. Firstly, if $\zeta$ is small, note that
\begin{equation}\label{Opt1TE}
	F(\zeta;s,e_{i})\approx \widetilde{F}(\zeta;s,e_{i}):=\Ex^{\widetilde{\Qx}}\left[\left. e^{I_{0,\zeta}} {\rm BS}\left(\zeta;se^{\zeta r' X_{\zeta}},(\sigma' X_{\zeta})^{2},0,K\right)\right|S_{0}=s,X_{0} = e_{i} \right],
\end{equation}
up to an error $O(\zeta)$. This is because $\int_{0}^{\zeta}r' X_{u}du=\zeta r' X_{\zeta}$ and $\int_{0}^{\zeta}(\sigma' X_{u})^{2}du=\zeta (\sigma' X_{\zeta})^{2}$ if there {are} no transitions of the process $(X_{t})$ during $[0,\zeta]$. Since the expression in (\ref{Opt1TE}) can be seen as a European claim of the form (\ref{BndPrc3}) with maturity $\zeta$ and payoff $\Xi(e_{j}):={\rm BS}\left(\zeta;se^{\zeta r_{j}},\sigma_{j}^{2},0,K\right)$, one can evaluate this ``first order approximation" using our Algorithm \ref{alg:FindOptPrice} {in Appendix \ref{Sec:PseudoCodesAll}.}

For a general maturity $\zeta$, we proceed as in (\ref{SmallTimeExp}). Concretely, for $\delta<\zeta$, we have the recursive relationship
\begin{align*}
	F(\zeta;s,e_{i})=\Ex^{\widetilde{\Qx}}\left[\left. e^{I_{0,\delta}}F\left(\zeta-\delta; S_\delta,X_{\delta}\right)\right|S_{0}=s,X_{0} = e_{i} \right]=\Ex^{\widetilde{\Qx}}\left[\left. e^{I_{0,\delta}}F\left(\zeta-\delta; s e^{\bar{b} \delta +\bar{\sigma}\sqrt{\delta} {\DMagenta W^{\Qx}_{1}}},X_{\delta}\right)\right|X_{0} = e_{i} \right],
\end{align*}
where  {now $\bar{\sigma}^{2}:=(\zeta-\delta)^{-1}\int_{\delta}^{\zeta}(\sigma' X_{u})^{2} du$ and  $\bar{b}:=(\zeta-\delta)^{-1}\int_{\zeta}^{\delta}\tilde{b}' X_{u} du$. As before, the last expression} is approximated by
\begin{align*}
	\hat{F}(\zeta;s,e_{i})&:=\Ex^{\widetilde{\Qx}}\left[\left. e^{I_{0,\delta}}F\left(\zeta-\delta; s e^{\delta \tilde{b}' X_{\delta} +\sigma' X_{\delta}\sqrt{\delta} {\DMagenta W^{\Qx}_{1}} },X_{\delta}\right)\right|X_{0} = e_{i} \right]\\
	&=\Ex^{\widetilde{\Qx}}\left[\left. e^{I_{0,\delta}}\int_{-\infty}^{\infty}F\left(\zeta-\delta; s e^{\delta \tilde{b}' X_{\delta} +\sqrt{\delta}\sigma' X_{\delta} z},X_{\delta}\right)\frac{1}{\sqrt{2\pi}}e^{-z^{2}/2}dz\right|X_{0} = e_{i} \right]{\DBlue .}
\end{align*}
In principle, for a {fixed} $s$, one can see the right-hand side above as the price of European claim of the form (\ref{BndPrc3}) with maturity $\delta$ and payoff
{$\Xi(e_{j}):=\Ex^{\widetilde{\Qx}}\Big[F\left(\zeta-\delta; s e^{\delta \tilde{b}_{j} +\sigma_{j}\sqrt{\delta} {\DMagenta W^{\Qx}_{1}}},e_{j}\right)\Big]$}. {But, since this approach would require to compute $F(\zeta-\delta; p,e_{j})$ for all $p$ and $e_{j}$, this would be computationally inefficient.} To resolve this issue, we restrict all possible initial prices $s$ to be in the lattice $\mathcal{L}_{\Delta,B}:=\{ s e^{i \Delta}: i\in \{-B,-B+1,\dots,B-1,B\}\}$ for a small $\Delta$ and a positive integer $B$. Then, we can approximate $F(\zeta; se^{i\Delta },e_{j})$ as follows:
\[
	F(\zeta;se^{i\Delta},e_{i})\approx\Ex^{\widetilde{\Qx}}\left[\left. e^{I_{0,\delta}}\,{\widetilde{\Xi}}_{i}(X_{\delta})\right|X_{0} = e_{i} \right],
\]
with
\begin{align*}
	{\widetilde{\Xi}_{i}(e_{j}):= \sum_{k=-B}^{B} F(\zeta-\delta; se^{k\Delta},e_{j})\int_{z_{k-1}^{i,j}}^{z_{k}^{i,j}} \frac{1}{\sqrt{2\pi}}e^{-z^{2}/2}dz},
\end{align*}
where
\[
	z_{k}^{i,j}:=\frac{(k-i+1/2)\Delta-\delta \tilde{b}_{j}}{\sigma_{j} \sqrt{\delta}}, \qquad(k=-B,-B+1,\dots,B-1), \qquad z_{-B-1}^{i,j}:=-\infty, \qquad z_{B}^{i,j}:=\infty.
\]
Note that the points $z_{k}^{i,j}$'s are chosen so that the midpoint $\bar{z}_{k}^{i,j}$ of the interval $[z_{k-1}^{i,j}, z_{k}^{i,j}]$ is such that $s \exp\{k\Delta\}=s \exp\{i\Delta+\delta\tilde{b}_{j}+\sqrt{\delta}\sigma_{j}\bar{z}_{k}^{i,j}\}$ and, hence, $\widetilde{\Xi}_{i}$ above is a Riemann-Stieltjes sum approximation of the payoff
\[
	\Xi_{i}(e_{j}):=\int_{-\infty}^{\infty}F\left(\zeta-\delta;  se^{i\Delta+\delta \tilde{b}_{j} +\sqrt{\delta} \sigma_{j}z},e_{j}\right)\frac{1}{\sqrt{2\pi}}e^{-z^{2}/2}dz.
\]

\section{Numerical Simulations} \label{sec:numerical}
The objective of this section is to assess the accuracy and computational speed of the {\DBlue different methods introduced in the previous section}. 
We {\DBlue first} describe the simulation scenario in Section \ref{sec:simscenario} and {\DBlue then} 
provide the numerical results in Section \ref{sec:numer}.

\subsection{Simulation Scenario}\label{sec:simscenario}
Throughout this part, we use the following parameter specifications borrowed from the parameter setup of  \cite{CapFigDyn}. The right panel also shows the loss rates {\Blue corresponding} to three distinct regimes, hereafter referred to as ``low'', ``middle'', and ``high'' default regime. The default intensities are based on the empirical result of \cite{Longstaff}\footnote{\cite{Longstaff} employed a three-state homogenous {\Awesome regime-switching} model to examine the effects of an array of financial and macro-economic variables in explaining variations in the realized default rates of the U.S. corporate bond market over the course of 150 years.}. {\Blue In agreement with} empirical market evidence, we {\YGreen have selected} the loss rates to be increasing in the credit riskiness of the regime.

\begin{table}[htp]
\begin{center}
\begin{tabular}{|c|c|c|c|}
  \hline
  $a_{i,j}^{\Qx}$ & 1 {(low)} & 2 {(middle)} & 3 {(high)} \\
  \hline
  \hline
  1 & -0.380313 & 0.33687 & 0.043443 \\
  \hline
  2 & 0.254397 & -0.254397 & 0 \\
  \hline
  3 & 0.208683 & 0.000006 & -0.208689 \\
  \hline
\end{tabular}
\hspace{2 cm}
\begin{tabular}{|c|c|c|}
  \hline
  & h & L \\
  \hline
  \hline
  1 & 0.741\% & 10\% \\
  \hline
  2 & 4.261\% & 40\% \\
  \hline
  3 & 11.137\% & 90\% \\
  \hline
\end{tabular}
\end{center}
\caption{{Left panel shows the generator of the Markov chain under the risk-neutral measure. The rows indicate the starting state, while the columns indicate the ending state of the chain. The right panel shows the default intensities as reported in \cite{Longstaff} as well as our loss rates given default associated to three regimes.}}
\label{tab:histgen}
\end{table}

\subsection{Numerical Results} \label{sec:numer}
We start {\Awesome by} illustrating the accuracy and computational speed of the {\DBlue method described in Section \ref{sec:BscMthd}} for computing the pre-default {\Blue price of a defaultable bond}. These bond {\DBlue prices}, denoted hereafter $\psi(t):=(\psi_{1}(t),\dots,\psi_{N}(t))'$, are given by the formula (\ref{BndPrc3}) with payoff function $\Xi(e_{i})\equiv 1$. {\DBlue Concretely, we define}
\begin{equation}\label{OrgDfnPsi}
	\psi_{i}(t):=\Ex^{\Qx}\left[\left. e^{-\int_{t}^{T}(r_{s} + h_{s}L_{s})ds} \right| X_{t} = e_{i} \right].
\end{equation}
We compute (\ref{OrgDfnPsi}) by our method and a standard numerical solution of the Feynman-Ka\v{c} representation (\ref{eq:bondspdeGen}) for the bond price (\ref{OrgDfnPsi}).
Table \ref{tab:Comparison} shows the time-$0$ bond prices for different maturities computed using our method and a Runge-Kutta type numerical solution\footnote{This solution was obtained using the MATLAB function ode45.} of the system (\ref{eq:bondspdeGen}) with terminal condition $\Psi_{i}(T)=1$ under the parameter setup of Table \ref{tab:histgen}. {Figure \ref{Pricing Methods Comparison} shows the time {$t$ bond price for maturity $T=1$ year (left panel), and $T=20$ years (right panel).}} {It is evident that our method} is highly accurate {even for long maturity bonds}. Furthermore, according to our {computational experiments}, our method is in {most cases} more efficient than {either solving the ODE system by the Runge-Kutta algorithm or computing} the exponential (\ref{eq:psiexpr}) using a Pad\'e {type} approximation. For the {sake of completeness}, Appendix \ref{Sec:SpeedCmp} compares the processor time for these three methods.
\begin{table}[ht]
\begin{center}
\begin{tabular}{|c|cc|cc|cc|}
  \hline
  \multicolumn{1}{|c|}{} & \multicolumn{2}{|c|}{$X_{0} = e_{1}$} & \multicolumn{2}{|c|}{$X_{0} = e_{2}$} & \multicolumn{2}{|c|}{$X_{0} = e_{3}$} \\
  \hline
  ${T}$ (yrs.)  & NM & ODE  & NM & ODE  & NM & ODE\\
  \hline
0.25 & 0.9921 & 0.9921 & 0.9884 & 0.9884 & 0.9686 & 0.9686 \\
0.50 & 0.9837 & 0.9837 & 0.9772 & 0.9772 & 0.9393 & 0.9393 \\
1.00 & 0.9659 & 0.9659 & 0.9555 & 0.9555 & 0.8864 & 0.8864 \\
2.00 & 0.9282 & 0.9281 & 0.9146 & 0.9146 & 0.7991 & 0.7990 \\
5.00 & 0.8140 & 0.8136 & 0.8029 & 0.8031 & 0.6274 & 0.6273 \\
10.0 & 0.6488 & 0.6484 & 0.6430 & 0.6431 & 0.4701 & 0.4701 \\
15.0 & 0.5170 & 0.5166 & 0.5131 & 0.5131 & 0.3691 & 0.3690 \\
20.0 & 0.4119 & 0.4116 & 0.4090 & 0.4090 & 0.2931 & 0.2930 \\
25.0 & 0.3282 & 0.3280 & 0.3259 & 0.3259 & 0.2334 & 0.2333 \\
30.0 & 0.2616 & 0.2613 & 0.2597 & 0.2597 & 0.1859 & 0.1858 \\
50.0 & 0.1055 & 0.1053 & 0.1047 & 0.1047 & 0.0750 & 0.0749 \\
   \hline
\end{tabular}
\caption{{{Time $t=0$ bond prices} for different time-to-maturities {$T$} using the ODE method and the {new method (NM)} with {parameters $M=2$ and $\delta=2.5$ years (see Algorithm \ref{alg:FindOptPrice} in Appendix \ref{Sec:PseudoCodesAll})}.}}
\label{tab:Comparison}
\end{center}
\end{table}

\begin{figure}[htp]
{\par \centering
    \includegraphics[width=7.9cm,height=9.2cm]{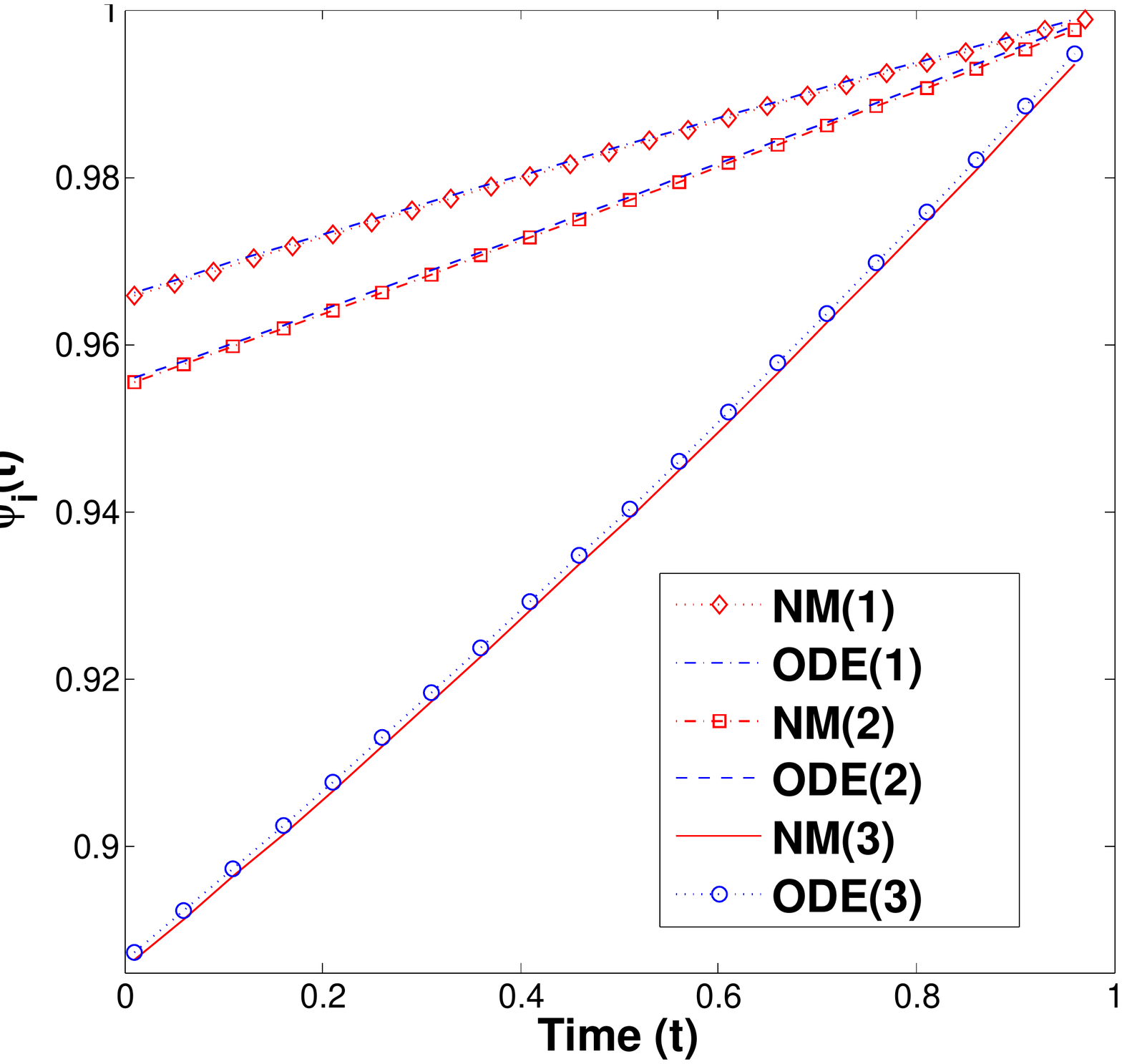}
     \includegraphics[width=7.9cm,height=9.2cm]{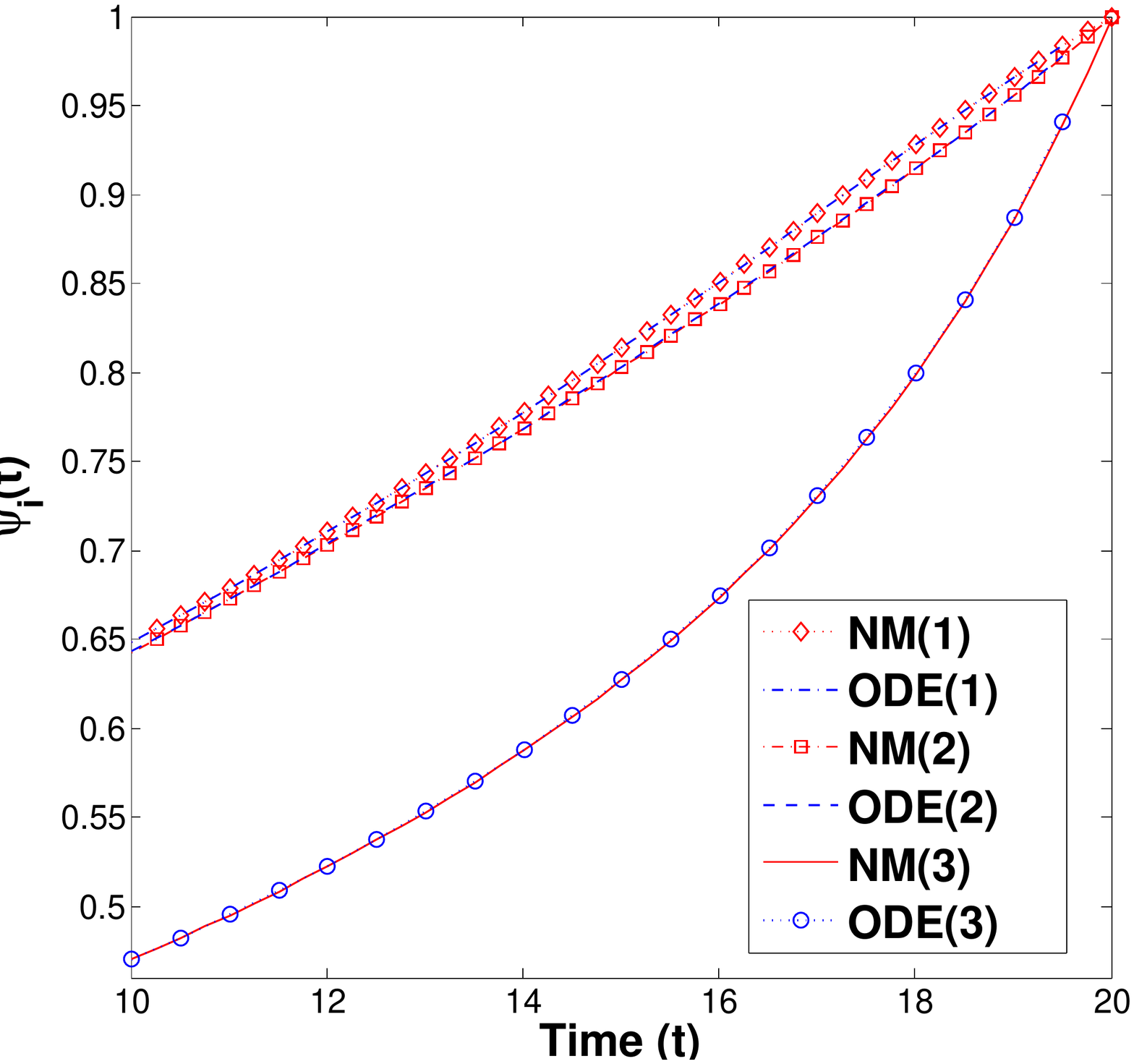}
   \par}\vspace{-1 cm}
    \caption{\label{Pricing Methods Comparison} Bond price comparison for a one-year and a {20-year} bond using the ODE method and the {new method (NM)} {parameters $M=2$ and $\delta=2.5$ years (see Algorithm \ref{alg:FindOptPrice} in Appendix \ref{Sec:PseudoCodesAll})}.}
\end{figure}

Next, we illustrate the power of the algorithm to compute prices of {\DBlue the knock-out digital and call options described at the end of Section \ref{sec:DcmpPrc}}.
Table \ref{tab:barrierComp} {reports the prices of {knock-out digital and call options} written on the volatility process {obtained by applying our methodology}, and compares them with the corresponding Monte-Carlo estimates}.  {The results show {\DBlue that our} method is highly accurate even for long maturity options.}
\begin{table}[ht]
\begin{center}
\begin{tabular}{|c|cc|cc|cc|cc|}
\hline
\multirow{4}{*}{} & \multicolumn{4}{|c|}{Knock-Out Digital Contracts} & \multicolumn{4}{|c|}{Knock-Out Barrier Call Options} \\
\multirow{4}{*}{} & \multicolumn{4}{|c|}{$\Xi(X_{T})\equiv 1$} & \multicolumn{4}{|c|}{$\Xi(X_{T}) := (\sigma_{X_{T}} - K)^{+}$} \\
& \multicolumn{4}{|c|}{$B=\sigma_{3}$} & \multicolumn{4}{|c|}{$B=\sigma_{3}$, $K= 0.075$, ${{\rm Units}=10^{-3}}$}\\
\hline
\multicolumn{1}{|c|}{} & \multicolumn{2}{|c|}{$X_{0} = e_{1}$} & \multicolumn{2}{|c|}{$X_{0} = e_{2}$}  & \multicolumn{2}{|c|}{$X_{0} = e_{1}$} & \multicolumn{2}{|c|}{$X_{0} = e_{2}$}\\
\hline
T & NM & MC & NM & MC & NM & MC & NM & MC\\
\hline
0.50 & 0.9719 & 0.9693 & 0.9525 & 0.9523 & 3.5081 & 3.5574 & 21.1632 & 21.0587 \\
1.0 & 0.9419 & 0.9455 & 0.9093 & 0.9100 & 5.8869 & 5.7910 & 18.2870 & 18.2697 \\
2.5 & 0.8482 & 0.8466 & 0.7980 & 0.7964 & 9.0631 & 9.0351 & 13.1068 & 13.0927 \\
5.0 & 0.7013 & 0.6986 & 0.6506 & 0.6481 & 9.1533 & 9.2635 & 9.3528 & 9.5317 \\
10.0 & 0.4732 & 0.4721 & 0.4373 & 0.4336 & 6.4919 & 6.3608 & 6.0298 & 5.9426 \\
15.0 & 0.3187 & 0.3201 & 0.2944 & 0.2955 & 4.3833 & 4.3335 & 4.0508 & 4.1597 \\
20.0 & 0.2146 & 0.2141 & 0.1983 & 0.1974 & 2.9521 & 2.9365 & 2.7274 & 2.8137 \\
25.0 & 0.1445 & 0.1445 & 0.1335 & 0.1341 & 1.9879 & 1.9577 & 1.8366 & 1.7792 \\
30.0 & 0.0973 & 0.0985 & 0.0899 & 0.0912 & 1.3386 & 1.3136 & 1.2367 & 1.2111 \\
35.0 & 0.0655 & 0.0676 & 0.0605 & 0.0608 & 0.9014 & 0.9038 & 0.8328 & 0.8228 \\
\hline
\end{tabular}
\caption{Knock-out digital and call {options} on the volatility process using the new method (NM) and {the} Monte Carlo method (MC). The call option prices are expressed on the {$10^{-3}$} scale and the results for $X_{0}=e_{3}$ have been omitted since they are knocked out at contract initiation. {Here, {$r = (0.01, 0.1, 0.3)'$} and $\sigma= (0.05, 0.1, 0.2)'$.}}
\label{tab:barrierComp}
\end{center}
\end{table}
We conclude the numerical analysis with an illustration of the algorithm performance on pricing vulnerable European type claims, which depend on both the diffusion and the Markov chain component. Figure \ref{ComparisonVulnerable} compares our novel method with a plain Monte Carlo method, given that the initial regime is 1 (left panel), and 3 (right panel) {(regime 2 is quite similar to regime 3)}. As seen there, the {new} method {significantly improves the quality of the approximation {compared} to the first order approximation (\ref{Opt1TE})}, especially for longer maturities. {For the sake of completeness, we have included the precise algorithm in Appendix \ref{Sec:PseudoCodesAll} (see Algorithm \ref{alg:FindVulnOptPrice} therein)}. Error analysis and further extensions of this method will be postponed for a future publication.

\begin{figure}[htp]
{\par \centering
    \includegraphics[width=8.0cm,height=9.2cm]{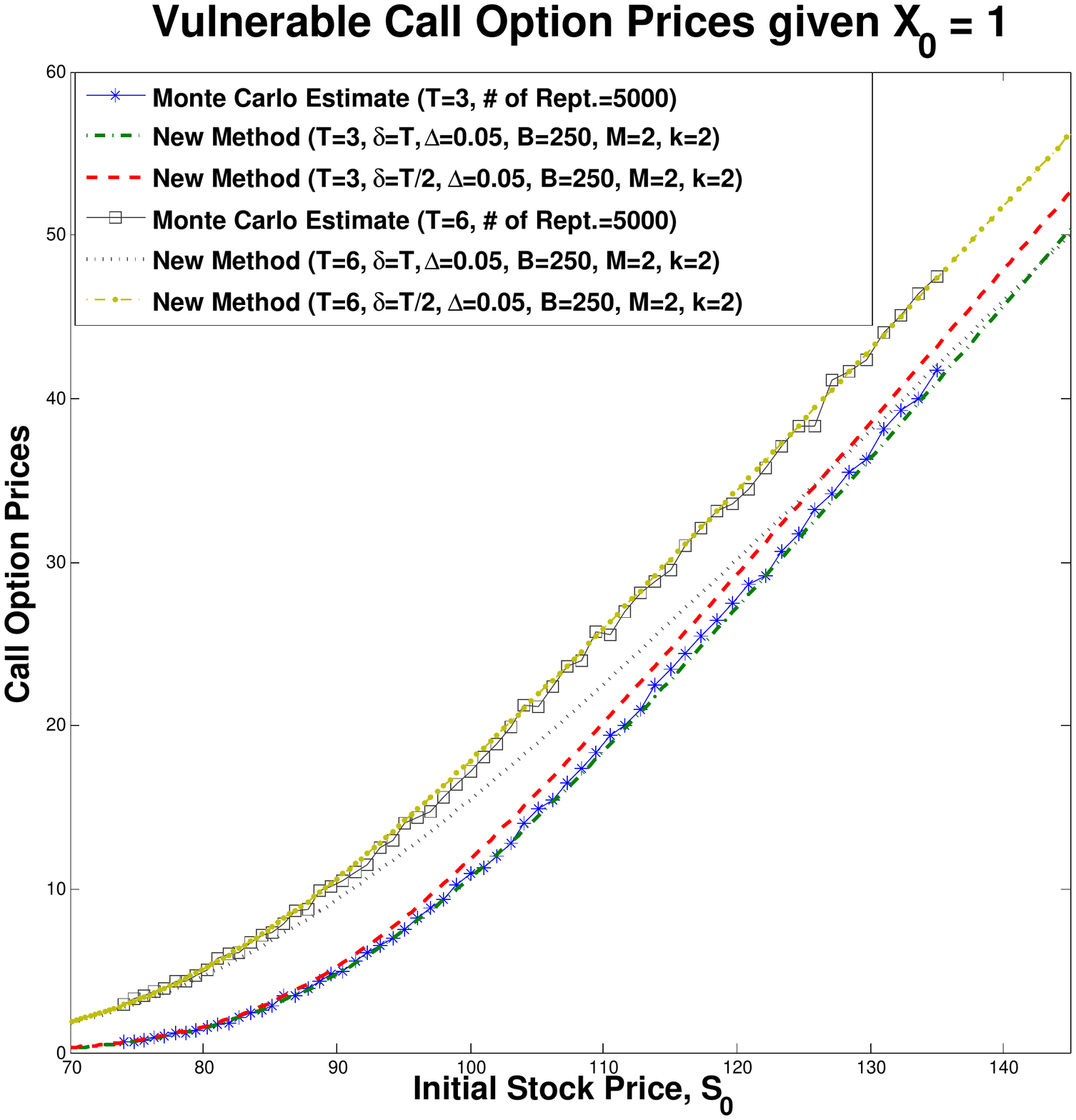}
 \hspace{1.0 cm}    \includegraphics[width=8.0cm,height=9.2cm]{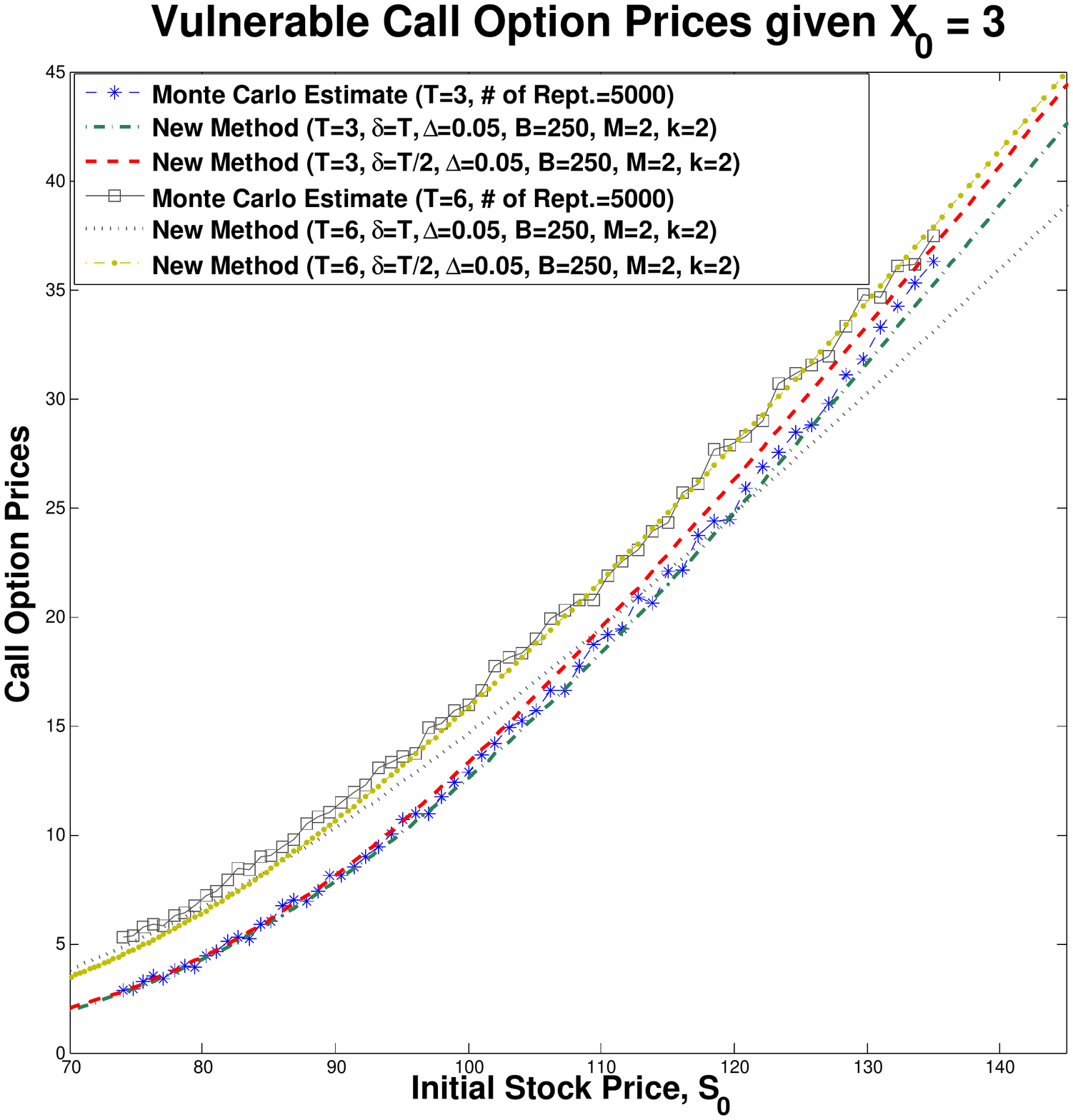}
   \par}\vspace{-1 cm}
    \caption{\label{ComparisonVulnerable} Comparison of vulnerable call option prices using Monte Carlo and the {new method}  of the Algorithm \ref{alg:FindVulnOptPrice} {in Appendix \ref{Sec:PseudoCodesAll}} for maturities of {$T=3$} years and {$T=6$} years. Above, {$k =2$} refers to the number of discretization points on the PriceClaim Method {\DBlue of Appendix \ref{Sec:PseudoCodesAll}} {(i.e. $\delta=T/k$ therein)}.}
\end{figure}

\section{Conclusions}\label{sec:conclusion}
{\DRed
We have rigorously analyzed pricing of vulnerable claims in {\Awesome regime-switching} markets, and provided {\Blue novel semimartingale} representations . The main theoretical tool underlying our results is a novel Poisson series representation of a generic claim price, whose payoff depends on the underlying Markov chain. Using this representation, we were able to prove differentiability of the pre-default price function under mild conditions on the Markov chain generator. This allowed us {\Awesome to provide} rigorous Feynman-Ka\v{c} and semimartingale representations of the price process of the vulnerable claim.
We then constructed a novel pricing methodology which makes use of the Poisson series representation, and expands the price of the vulnerable claim in terms of the Laplace transform of the symmetric Dirichlet distribution. We provided a detailed error analysis of our proposed algorithm under different sets of assumptions relating to how {\Blue each conditional price component} may be approximated. {\Blue One of the novel features of our method and error analysis is that it also applies to time-dependent Markov-chain generators}. We further demonstrated the robustness of our methodology by showing that it can be used to price path dependent claims, termed self-decomposable, as well as vulnerable European-type options, whose payoff depends on both the diffusion and the Markov-chain component. We numerically demonstrated the algorithm on defaultable bonds, barrier options on the stock volatility, as well as vulnerable European call option prices. In all these cases, we have shown that it achieves a high level of accuracy, and, at the same time, it is computationally fast when compared to standard PDE or Monte-Carlo based methods.
}

{\Blue
\subsection*{Acknowledgments}
The second author's research was partially supported by a grant from the US National Science Foundation (DMS-0906919).  {\DViolet The third author's research was {\Awesome partially} supported by a Purdue Research Foundation Ph.D. student grant.} The authors gratefully acknowledge the constructive and insightful comments provided by two anonymous {\DViolet referees}, which {\DViolet significantly} contributed to improve the quality of this manuscript.}

\appendix

\section{Proofs}\label{NmrMthdApdx}

\begin{proof}[{\bf Proof of Lemma \ref{LmNdFD}}]\hfill

\noindent
By the definition of $\Qx$ and the fact that $\eta$ is a ${\Gx}$-martingale (see the {paragraph} before (\ref{DfnDnsty})),  it suffices to prove that $\eta_{t} \xi_{t}^{\Qx}$ is a ${\Gx}$-martingale under $\Px$. From It\^o's formula and the definition of $\eta$ in (\ref{DfnDnsty}), we have the process
	\begin{align*}
		\eta_{t} {\xi^{\Qx}_t}={\xi^{\Qx}_0}+\int_{0}^{t} \eta_{s^{-}} {d\xi^{\Qx}_s} +
		\int_{0}^{t} \xi_{s^{-}}^{\Qx} d\eta_{s}
		+\sum_{s\leq{}t} \Delta \xi_{s}^{\Qx} \Delta \eta_{s}.
	\end{align*}
	From (\ref{MrtRprDftPrcP}), (\ref{DfnDnsty}), and (\ref{MrtRprDftPrcQ}), {$(\eta_{t} {\xi^{\Qx}_t})_{t}$ can be written {as}}
	\begin{align*}
		{\xi^{\Qx}_0} +\int_{0}^{t} \eta_{s^{-}} {\YGreen d\xi^{\Qx}_s} +
		\int_{0}^{t} \xi_{s^{-}}^{\Qx} d\eta_{s}+\sum_{0<s\leq{}t} \sum_{k,l=1}^{N} \eta_{s^{-}} \kappa_{k,l}(u) \Delta H(s)
		\Delta H_{s}^{k,l},
	\end{align*}
	where $H_{t}^{k,l}:=\sum_{0<s\leq{}t}\idc_{\{X_{s^{-}}=k\}} \idc_{\{X_{s}=l\}}$. Since the first two terms on the right-hand side of the previous equality are (local) martingales under $\Px$, it remains to show that the last term vanishes. But, given that $\Delta H_{s}\neq{}0$ at $s=\tau$, the summation in the last term above will be $0$ provided that
	\(
		{\Delta X_{\tau}=0,}
	\)
	a.s.
	In order to show this, let us recall that by definition $X$ has no fixed-jump times; i.e. $\Px(\Delta X_{t}\neq{}0)=0$ for any fixed time $t>0$. Also, using the definition of $\tau$ given in Eq.~(\ref{eq:taudef}), $\tau=\inf\{t\geq{}0: \int_{0}^{t} h(X_{s})ds\geq{}{\Blue {\YGreen \Theta}}\}$, where {\Blue ${\YGreen \Theta}$} is an exponential random variable independent of $X$. Then, conditioning on $X$,
	\(
		{\Px(\Delta X_{\tau}\neq{}0)=\Ex\left\{\Ex\left[\left.{\bf 1}_{\{\Delta X_{\tau}\neq{}0\}}\right|X_{s},s\geq{}0\right]\right\}}.
	\)
	Denoting $0<\tau_{1}<\tau_{2}<\dots$ the transition times of the Markov chain $X$, {$\Ex\left[\left.{\bf 1}_{\{\Delta X_{\tau}\neq{}0\}}\right|X_{s},s\geq{}0\right]$ is given by}
	\begin{align*}
		\Ex\left[\left.\sum_{i=1}^{\infty}{\bf 1}_{\{\tau=\tau_{i}\}}\right|X_{s},s\geq{}0\right]= \Ex\left[\left.\sum_{i=1}^{\infty}{\bf 1}_{\{\int_{0}^{\tau_{i}} h(X_{s})ds={\Blue {\YGreen \Theta}}\}}\right|X_{s},s\geq{}0\right]=0,
	\end{align*}
	where the last equality {follows} from the independence of $X$ and {\Blue ${\YGreen \Theta}$}, and the fact that {\Blue ${\YGreen \Theta}$} is a continuous random variable.
\end{proof}

\begin{proof}[{\bf Proof of {\DRed Theorem \ref{Lm:DfPsi}}}]\hfill

\noindent
For simplicity we omit the dependence on the claim's payoff $\Xi$ and maturity $T$ in (\ref{BndPrc3}) and just write $\Psi_{i}(t)$. Recall that $\Psi(t):=(\Psi_{1}(t),\dots,\Psi_{N}(t))$ is such that
	\begin{equation}
		\gamma(t;T) = \Ex^{\mathbb{Q}}\left[\Xi (X_{T})e^{-\int_t^T (r_s + h_s L_s) ds} | X_t \right]=\left<\Psi(t),X_{t}\right>.
	\end{equation}
But also, changing into the probability measure $\widetilde{\Qx}$, we can write
\[
	\gamma(t;T) = \Ex^{{\Qx}}\left[\left. \Xi (X_{T})e^{-\int_t^T (r_s + h_s L_s) ds} \right| \mathcal{F}_t \right]
	=\Ex^{\widetilde{\Qx}}\left[\left. \Xi (X_{T})e^{-\int_t^T (r_s + h_s L_s) ds} {\Blue \frac{\tilde{\eta}_{_{t}}}{\tilde\eta_{_{T}}}}\right| \mathcal{F}_t \right],
\]	
and, hence, we have the following representation for $\Psi_{i}$:
\begin{equation}\label{Eq:KELRp}
	\Psi_{i}(t)  =\Ex^{\widetilde{\Qx}}\left[\left. e^{-\int_t^T (r_s + h_s L_s) ds} {\Blue \frac{\tilde{\eta}_{_{t}}}{\tilde\eta_{_{T}}}}\right| X_{t}=e_{i}\right].
\end{equation}
The solution of (\ref{DfnDnstyHom}) can be written as
\[
	{\tilde{\eta}_{t}:=e^{-\int_{0}^{t} \sum_{i,j}a_{i,j}^{\Qx}(s)\tilde{\kappa}_{i,j}(s)H^{i}_{s}ds
	+\sum_{0<s\leq{}t} \log\left(1+\sum_{i,j}\tilde\kappa_{i,j}(s)\Delta H^{i,j}_{s}\right)}};
\]
see pp. 334-335 in \cite{BielCrep08}.
Let $\widetilde{K}(t)=[\widetilde{K}_{i,j}(t)]_{i,j}$ and $\tilde{r}(t):={\Blue (\tilde{r}_{1}(t),\dots,\tilde{r}_{N}(t))'}$ be defined by
\[
	\widetilde{K}_{i,j}(t):=\log\left(1+\tilde\kappa_{i,j}(t)\right){\bf 1}_{i\neq j}=-\log\left((N-1)a_{i,j}^{\Qx}(t)\right){\bf 1}_{i\neq j},\qquad
	 \tilde{r}_{i}(t):=r_{i}+h_{i}L_{i}{\Blue -}\sum_{j=1}^{N}a_{i,j}^{\Qx}(t)\tilde\kappa_{i,j}(t).
\]
Then, since by the definition (\ref{JmpTrnPrc1}), $\Delta H_{s}^{i,j}= 1$ whenever there is a jump from the state $e_{i}$ to the state $e_{j}$ (i.e. $X_{s^{-}}=e_{i}$ and $X_{s}=e_{j}$) and $\Delta H_{s}^{i,j}= 0$, otherwise, we have
\[
	\sum_{s\leq{}t} \log\bigg(1+\sum_{i,j}\tilde\kappa_{i,j}(s)\Delta H^{i,j}_{s}\bigg)=\sum_{s\leq{}t} \sum_{i,j}\log\left(1+\tilde\kappa_{i,j}(s)\right)\Delta H^{i,j}_{s}= \sum_{s\in(0,t]:\Delta X_{s}\neq{}0}  X_{s^{-}}'\widetilde{K} (s) X_{s}.
\]
Similarly, by the definition (\ref{JmpTrnPrc1}),
\[
	\int_{0}^{t} \sum_{i,j}a_{i,j}^{\Qx}(s)\tilde{\kappa}_{i,j}(s)H^{i}_{s}ds=\int_{0}^{t} v(s)X_{s}ds,
\]
where $v=(v_{1},\dots,v_{N})'$ with $v_{i}(s):=\sum_{j=1}^{N}a_{i,j}^{\Qx}(s)\tilde{\kappa}_{i,j}(s)$. Then, since $r_{s}+h_{s}L_{s}=\hat{r}' X_{s}$ with $\hat{r}=(r_{1}+h_{1}L_{1},\dots, r_{N}+h_{N}L_{N})$, we obtain from (\ref{Eq:KELRp}) that
\[
	\Psi_{i}(t)=\Ex^{\widetilde{\Qx}}\left[\left. \exp\left\{ -\int_t^T \tilde{r}(s)' X_{s} ds
	-\sum_{s\in(t,T]:\Delta X_{s}\neq{}0}  X_{s^{-}}'\widetilde{K} (s) X_{s}\right\}\right| X_{t}=e_{i}\right],
\]
where $X'$ denotes the transpose of $X$.
\end{proof}

\begin{proof}[{\bf Proof of Lemma \ref{Cor:DiffPrice}}]\hfill

\noindent
Again, for simplicity we omit the dependence on the claim's payoff $\Xi$ and maturity $T$ in the functions $\Psi_{i}[\Xi](t;T)$ and $\Phi_{i,m}[\Xi](t;T)$.
In light of the series representation (\ref{CmpPsii}), we see that it suffices to show the function $\Phi_{i,m}(\zeta)$
is continuously differentiable in $\zeta\in(0,T)$ for each $m\geq{}0$, and that there exists a sequence $\{K_{m}\}_{m\geq{}0}$ of reals such that
\[
	{\rm (i)}\;\sup_{0\leq \zeta\leq  T}|\Phi_{m}(\zeta)|\leq{} K_{m},\quad
	{\rm (ii)}\;\sup_{0\leq \zeta\leq T}|\Phi_{m}'(\zeta)|\leq{} K_{m},\quad
	{\rm (iii)}\;\sum_{m=0}^{\infty} e^{-\zeta}\frac{\zeta^{m}}{m!} K_{m}<\infty.
\]
We now show that (i)-(iii) are satisfied provided the function $\widetilde{K}$ and $\tilde{r}$ defined in (\ref{Eq:DfnTildeKTilder}) are such that
\begin{equation}\label{Eq:CndNdDiff}
	\sup_{s\in[0,T]} |\tilde{r}_{i}(s)|<\infty,\quad
	\sup_{s\in[0,T]} |\widetilde{K}_{i,j}(s)|<\infty,\quad\text{ and }\quad
	\sup_{s\in[0,T]} |\widetilde{K}'_{i,j}(s)|<\infty.
\end{equation}
The conditions (\ref{Eq:CndNdDiff}) directly follow from (\ref{NdCndDf1}) and (\ref{NdCndDf1b}).
Denoting $\mathcal{P}_{m,i}(\zeta)$ the random function inside the expectation $\Ex^{\widetilde{\Qx}}_{i}$ in (\ref{AuxDfn1b}), one can check that
\begin{equation}\label{Dmmy1}
	\sup_{0\leq\zeta\leq{}T}|\mathcal{P}_{m,i}(\zeta)|\leq{}{\Awesome \Xi(e_i)} \cdot e^{m T\max_{i}\sup_{0\leq{}\zeta\leq T} |{\Blue \tilde{r}_{i}(\zeta)'}|
	+m \max_{i,j}\sup_{0\leq{}\zeta\leq T} |\widetilde{K}_{i,j}(\zeta)|}={\Awesome \Xi(e_i)} A^{m},
\end{equation}
for a constant $A<\infty$.
Also, $\mathcal{P}_{m,i}(\zeta)$ is continuously differentiable and
\begin{align*}
	 \mathcal{P}_{m,i}'(\zeta)&=\mathcal{P}_{m,i}(\zeta)\left\{\sum_{n=0}^{m}\left[\bar{U}_{(n+1)}{\Blue \tilde{r}(T-\zeta\bar{U}_{(n+1)})'\widetilde{X}_{n}}
	 -\bar{U}_{(n)}{\Blue \tilde{r}(T-\zeta\bar{U}_{(n)})'}{\Blue \widetilde{X}_{n}}\right] - \sum_{n=1}^{m}{\Blue \widetilde{X}'_{n-1}\widetilde{K}' (T-\zeta\bar{U}_{(n)}) \widetilde{X}_{n}}\bar{U}_{(n)} \right\},
\end{align*}
where $\bar{U}_{(n)}:=1-U_{(n)}$. In particular, there exists a constant $B<\infty$ such that
\begin{equation}\label{Dmmy2}
	\sup_{0\leq\zeta\leq{}T}|\mathcal{P}'_{m}(\zeta)|\leq{}  B m A^{m}.
\end{equation}
By the formal definition of the derivative $\Phi_{m,i}'(\zeta)$ and the {\YGreen Dominated Convergence Theorem},  one can check that (\ref{Dmmy1})-(\ref{Dmmy2}) will suffice for (i)-(iii).

\end{proof}

\begin{proof}[{\bf Proof of Theorem \ref{Th:FKRNDyn}}]\hfill

\noindent
The following three parts {\Blue show} each of the statements of Theorem \ref{Th:FKRNDyn}:

\smallskip
\noindent
\textbf{(1)}
Consider the {$({\Fx},\Qx)$}-{\DMagenta local} martingale
\[
	Z(t):= \Ex^{\mathbb{Q}}\left[\Xi(X_{T})e^{-\int_0^T (r_s + h_s L_s) ds} |\mathcal{F}_t \right].
\]
Note that
\[
	Z(t) = e^{-\int_0^t (r_s + h_s L_s) ds}\gamma(t)=e^{-\int_0^t (r_s + h_s L_s) ds} \left<\Psi(t), X_t\right>,
\]
and, thus, by It\^o's formula and the decomposition (\ref{eq:MCsemimb}), we get
\begin{align}\nonumber
dZ(t) &= -(r_t + h_t L_t) e^{-\int_0^t (r_u + h_u L_u) du} \left<{\Psi(t)}, X_t\right> dt  + e^{-\int_0^t (r_u + h_u L_u) du} \Big(\Big< \frac{d\Psi(t)}{dt}, X_t\Big> + \left<\Psi(t), (A^{\Qx})'_{t} X_t\right>\Big) dt\\
&\quad + e^{-\int_0^t (r_u + h_u L_u) du} {\left<{\Psi(t)}, dM^{\Qx}(t)\right>},\label{Eq:DynForZ}
\end{align}
where we have used the fact that $\Psi(t)$ is differentiable in $t$ by virtue of Lemma \ref{Cor:DiffPrice}. Since $Z(t)$ is a {$({\Fx},\Qx)$-(local)} martingale, its drift term must be zero. When $X_{t}=e_{i}$, such condition translates into the {\DMagenta coupled system of differential equations for the price function $\Psi_{i}$ given in Eq.~\eqref{eq:bondspdeGen}}.

\smallskip
\noindent
\textbf{(2)}
The following dynamics for $Z$ also follows from (\ref{Eq:DynForZ}) and the martingale property of $\{Z_{t}\}_{t\geq{}0}$:
\begin{equation*}
dZ(t) =  e^{-\int_0^t (r_u + h_u L_u) du} {\left<{\Psi(t)}, dM^{\Qx}(t)\right>}.
\end{equation*}
In particular, the {price process} $\gamma(t)$ follows the dynamics
\begin{align}\nonumber
	d\gamma(t)&=d\left(e^{\int_0^t (r_s + h_s L_s) ds}Z(t)\right)\\
	&=(r_{t}+h_{t}L_{t})e^{\int_0^t (r_s + h_s L_s) ds}Z(t) dt+e^{\int_0^t (r_s + h_s L_s) ds} d Z(t) \nonumber\\
	&=(r_{t}+h_{t}L_{t})e^{\int_0^t (r_s + h_s L_s) ds}Z(t) dt+\left<{\Psi(t)}, dM^{\Qx}(t)\right>.
	\label{Eq:PrgDyngamma}
\end{align}
Next, denoting $\tilde{H}(t):=1-H(t)={\bf 1}_{\tau>t}$ and applying It\^o's formula,
\begin{equation}\label{Eq:Dyngamma1}
	d\Gamma(t;T)=d(\tilde{H}(t)\gamma(t))=\tilde{H}(t^{-}) d\gamma(t) + \gamma(t^{-}) d\tilde{H}(t) + \Delta \gamma(t) \Delta \tilde{H}(t).
\end{equation}
By the continuity of $\Psi(t)$,
\[
	 \Delta \gamma(t)\Delta \tilde{H}_{t}=\left<\Psi(t),\Delta X_{t}\right>\Delta \tilde{H}_{t}=\left<\Psi(t),\Delta X_{\tau}\right>=0, \quad \text{a.s.},
\]
where the last equality hold true in light of the construction (\ref{eq:taudef}) and the independence of $X$ and the exponential variable ${\YGreen \Theta}$ there {\Blue (see also the proof of Lemma \ref{LmNdFD} above)}. Combining (\ref{Eq:PrgDyngamma}) and (\ref{Eq:Dyngamma1}),
\begin{align*}
	d\,\Gamma(t;T)&=\tilde{H}(t^{-})(r_{t}+h_{t}L_{t})e^{\int_0^t (r_s + h_s L_s) ds}Z(t) dt+\tilde{H}(t^{-})\left<{\Psi(t)}, dM^{\Qx}(t)\right>+ \gamma(t^{-})d \tilde{H}(t)\\
	&=(r_{t}+h_{t}L_{t})\Gamma(t;T) dt+\tilde{H}(t^{-})\left<{\Psi(t)}, dM^{\Qx}(t)\right>- \gamma(t^{-})d H(t)
\end{align*}
By the Doob-Meyer decomposition of Lemma \ref{LmNdFD}, we have $dH(t) = d\xi_t^{\Qx} + (1-H(t^{-})) h_t dt$ and, thus,
\begin{align*}
	d\,\Gamma(t;T)=\left([r_{t}+h_{t}L_{t}]\Gamma(t;T)-\gamma(t^{-})h_{t}[1-H(t^-)]\right) dt+\tilde{H}(t^{-})\left<{\Psi(t)}, dM^{\Qx}(t)\right> -\gamma(t^{-}) {d\xi_t^{\Qx}}
\end{align*}
The last step follows from the fact that on the event $\tau>t$, we have $\Gamma(t^{-};T)=\gamma(t^{-};T)$, by definition of $\Gamma(t,T)$, and also $\gamma(t^{-}) =\left<\Psi(t), X_{t^{-}}\right>$. Therefore, we can write the risk-neutral dynamics of prices process $\Gamma(t;T)$ as
\begin{eqnarray}\label{DRNM2}
\frac{d\Gamma(t;T)}{\Gamma(t^{-};T)} &=& (r_t + h_t (L_t-1)) dt + \frac{{\left<{\Psi(t)}, dM^{\Qx}(t)\right>}}{{\left<\Psi(t), X_{t^{-}}\right>}} - d\xi_t^{\Qx},\qquad t<\tau,
\end{eqnarray}
which concludes the second statement of the theorem.

\smallskip
\noindent
\textbf{(3)}
The price dynamics of the vulnerable claim under the real-world measure follows directly from plugging Eq. (\ref{RMM0}) into (\ref{DRNM2}) to get:
\begin{align*}
\frac{d\Gamma(t,T)}{{\Gamma(t^{-},T)}} = \left(r_t + h_t (L_t-1)+
\frac{\left<\Psi(t),(A'(t)-(A^{\mathbb{Q}})'(t))X_{t}\right>}{
\left<\Psi(t),X_{t}\right>} \right) dt+ \frac{\left<\Psi(t), dM^{\Px}(t)\right>}{{\left<\Psi(t), X_{t^{-}}\right>}}-d\xi^{\Px}_{t},
\end{align*}
where we also used that $\xi^{\Px}=\xi^{\Qx}$ (see Lemma \ref{LmNdFD}).
Finally,  we get the dynamics (\ref{DRWP0}) since
\begin{align}\label{AuxQnt3}
	 \frac{\left<\Psi(t),(A'(t)-(A^{\mathbb{Q}})'(t))X_{t}\right>}{\left<\Psi(t),X_{t}\right>}&=
	 \frac{\left<(A(t)-A^{\mathbb{Q}}(t))\Psi(t),X_{t}\right>}{\left<\Psi(t),X_{t}\right>}.
\end{align}
When $X_{t}=e_{i}$, the previous quantity simplifies as
\[
	 D_{i}(t):=\sum_{j=1}^{n}(a_{i,j}(t)-a_{i,j}^{\Qx}(t))\frac{\Psi_{j}(t)}{\Psi_{i}(t)},
\]
and (\ref{AuxQnt3}) can be written as $\left<(D_{1}(t),\dots,D_{N}(t))',X_{t}\right>$.

\end{proof}

\begin{proof}[{\bf Proof of Theorem \ref{Th:FKRNDynCall}}]\hfill

\noindent
{\DMagenta The following three parts {\Blue show} each of the statements of Theorem \ref{Th:FKRNDynCall}:

\smallskip
\noindent
\textbf{(1)}
Consider the {$({\Fx},\Qx)$}-{\DMagenta local} martingale
\[
	Z(t,S_t):= \Ex^{\mathbb{Q}}\left[\varrho(S_{T})e^{-\int_0^T (r_s + h_s L_s) ds} |\mathcal{F}_t \right].
\]
Note that
\[
	Z(t,S_t) = e^{-\int_0^t (r_s + h_s L_s) ds} {\Awesome \gamma(t;S_t)} = e^{-\int_0^t (r_s + h_s L_s) ds} \left<\Pi(t;S_t),X_t \right>,
\]
where we used the notation $\gamma(t;S_{t})=\Ex^{\mathbb{Q}}\left[\varrho(S_{T})e^{-\int_t^T (r_s + h_s L_s) ds} |\mathcal{F}_t \right]$.
Thus, by It\^o's formula, we get
\begin{align}\nonumber
dZ(t,S_t) &= -(r_t + h_t L_t) e^{-\int_0^t (r_u + h_u L_u) du} \left<\Pi(t;S_t),X_t \right> dt + e^{-\int_0^t (r_u + h_u L_u) du} d \left<\Pi(t;S_t),X_t \right>.
\label{Eq:DynForZCall}
\end{align}
Under the stated differentiability assumptions, we have that
\begin{equation}
d \left<\Pi(t;S_t),X_t \right> = \mathcal{L} \left<\Pi(t;S_t),X_t \right> dt + d\mathcal{M}_{t},
\end{equation}
where $\mathcal{L}$ is the generator {\Blue of $(t, S_t, C_t)$}, and {\Blue $\{\mathcal{M}_t\}_{t\geq{}0}$ is} the local martingale component of $(t, S_t, C_t)$, under the risk neutral measure $\Qx$. These have {\Blue been} given in a more general context in \cite{CapFigDyn}, Appendix A. In our specific setting, they reduce to
\begin{equation}\label{eq:gencall}
\mathcal{L} \Pi_i(t;s) = \frac{\pa \Pi_i(t;s)}{\pa t} + r_i s \frac{\pa \Pi_i(t;s)}{\pa s} + \frac{\sigma_i^2 s^2}{2} \frac{\pa^2 \Pi_i(t;s)}{\pa^2 s} + \sum_{j \neq i} a_{i,j}^{\Qx}(t) \left( \Pi_j(t;s) - \Pi_i(t;s)  \right),
\end{equation}
and
\begin{equation}
d \mathcal{M}_{t} = \sum_{i=1}^N \bigg\{ \sum_{j \neq i} \left[ \Pi_j(t;S_{t-}) - \Pi_i(t;S_{t-}) \right] dM_t^{i,j} + \sigma_i S_t \frac{\pa \Pi_i(t;S_{t-})}{\pa s} H_t^i dW^{\Qx}_t \bigg\}.
\label{eq:locmart}
\end{equation}
{\Awesome where  $H_t^i$ and $M_t^{i,j}$ have been defined in Eqs.~\eqref{JmpTrnPrc1} and \eqref{CrsMrt}, respectively.} 
Using the operator $\mathcal{L}$ and local martingale $\mathcal{M}_{t}$ given by Eq.~\eqref{eq:gencall} and \eqref{eq:locmart}, respectively, we obtain
\begin{align}\nonumber
dZ(t,S_t) &= -(r_t + h_t L_t) e^{-\int_0^t (r_u + h_u L_u) du} \left<\Pi(t;S_t),X_t\right> {dt} + e^{-\int_0^t (r_u + h_u L_u) du} \left[(\mathcal{L} \left<\Pi(t;S_t),X_t \right>) dt +  d \mathcal{M}_{t} \right] \\
 \nonumber &= e^{-\int_0^t (r_u + h_u L_u) du} \bigg[-(r_t + h_t L_t) \left<\Pi(t;S_t),X_t \right> + \frac{\pa \left<\Pi(t;S_t),X_t \right>}{\pa t} + r_t S_t \frac{\pa \left<\Pi(t;S_t),X_t \right>}{\pa s}  \\
 & \quad +\; \frac{\sigma^2_{t} S_t^2}{2} \frac{\pa^2 \left<\Pi(t;S_t),X_t \right>}{\pa^2 s} + \sum_{j \neq C_{t-}} a_{C_{t},j}^{\Qx}(t) \left( \Pi_{j}(t;S_t) -\Pi_{C_{t-}}(t;S_t) \right) \bigg] dt + e^{-\int_0^t (r_u + h_u L_u) du} d \mathcal{M}_{t}.
\label{Eq:DynForZCallrew}
\end{align}
As {\Blue $\{Z(t,S_t)\}_{t\geq{}0}$} is a {\DMagenta{$({\Fx},\Qx)$}-local martingale}, its drift must vanish. When $X_t = e_i$, this translates into the coupled system of partial differential equations given by Eq.~\eqref{eq:callpdeGen}.

\smallskip
\noindent
\textbf{(2)}
In light of the martingale property of $Z(t,S_t)$, we have that its dynamics must be
\begin{equation*}
dZ(t,S_t) = e^{-\int_0^t (r_u + h_u L_u) du} d  \mathcal{M}_{t}.
\end{equation*}
Using results in Chapter 11 of \cite{bielecki01}, we can re-write the first term of {$d\mathcal{M}_{t}$} as
\begin{align}
{\sum_{i=1}^N \sum_{j \neq i} \left[ \Pi_j(t;S_{t}) - {\Pi_i(t;S_{t})} \right] dM_t^{i,j} = \left<\Pi(t;S_{t}), dM_t^{\Qx} \right>},
\label{eq:locmartrew}
\end{align}
{\Awesome Moreover, from the definition of $H_t^i = \idc_{\{X_{t}=e_i\}}$, given in Eq.~\eqref{JmpTrnPrc1}, we have that the second term of {$d\mathcal{M}_{t}$} can be written as 
\begin{align}
\sum_{i=1}^N  \sigma_i S_t \frac{\pa \Pi_i(t;S_{t})}{\pa s} H_t^i dW^{\Qx}_t  &= \sigma_t S_t \frac{\pa \left<\Pi(t;S_{t}),X_t \right>}{\pa s} dW_t^{\Qx},
\end{align}
and, thus, altogether}
\begin{equation}
dZ(t,S_t) = e^{-\int_0^t (r_u + h_u L_u) du} \left( \left<\Pi(t;S_{t}), dM_t^{\Qx} \right> + {\Awesome \sigma_t} S_t \frac{\pa \left<\Pi(t;{\Awesome S_{t}}),X_t \right>}{\pa s} dW_t^{\Qx}  \right),
\end{equation}
In particular, the process ${\Awesome \gamma(t;S_t)} = e^{\int_0^t (r_u + h_u L_u) du}Z(t,S_t)$ follows the dynamics
\begin{equation*}
d{\Awesome \gamma(t;S_t)}= (r_t + h_t L_t)  e^{\int_0^t (r_u + h_u L_u) du} Z(t,S_t) dt + \left<\Pi(t;S_{t}), dM_t^{\Qx} \right> + {\Awesome \sigma_t}  S_t \frac{\pa \left<\Pi(t;{\Awesome S_{t}}),X_t \right>}{\pa s} dW^{\Qx}_t.
\end{equation*}
Denoting $\tilde{H}(t):=1- H(t) = {\bf 1}_{\tau>t}$, we have that the vulnerable claim {\Awesome price process} $\Gamma(t,T;S_t) := \tilde{H}(t) {\gamma(t;S_t)}$ has dynamics
\begin{equation}\label{Eq:Dyngamma1call}
	d\Gamma(t,T;S_t)=d(\tilde{H}(t) {\gamma(t;S_t)})=\tilde{H}(t^-) d{\Awesome \gamma(t;S_t)} + {\Awesome \gamma(t;S_{t})} d\tilde{H}(t) + \Delta {\Awesome \gamma(t;S_t)} \Delta \tilde{H}(t).
\end{equation}
where, {using the hypothesis that $\Pi(t;S_t)$ is continuously differentiable in $t$, and twice continuously differentiable in $S_t$,} we have
\[
	 \Delta {\Awesome \gamma(t;S_t)} \Delta \tilde{H}_{t}=\left<\Pi(t;S_t),\Delta X_{t}\right>\Delta \tilde{H}_{t}=\left<\Pi(t;S_t),\Delta X_{\tau}\right>=0, \quad \text{a.s.}
\]
{with the last equality being true in light of the construction (\ref{eq:taudef}) and the independence of $X$ and the exponential variable ${\YGreen \Theta}$.}
By the Doob-Meyer decomposition of Lemma \ref{LmNdFD}, we have $dH(t) = d\xi_t^{\Qx} + (1-H(t^{-})) h_t dt$ and, thus,
\begin{align}\label{Eq:Dyngamma1callrew}
\nonumber d\Gamma(t,T;S_t) &= \left( (r_t + h_t L_t) \Gamma(t,T;S_t) - {\Awesome \gamma(t;S_{t})} h_t {\Awesome \tilde{H}(t^-)} \right) dt + \tilde{H}({\Awesome{t^-}}) {\Awesome \sigma_t} S_t \frac{\pa \left<\Pi(t;S_t),X_t\right>}{\pa s} dW^{\Qx}_t \\
& \quad + \tilde{H}(t)  \left<{\Awesome \Pi(t;S_{t})}, dM_t^{\Qx} \right> - {\Awesome \gamma(t;S_{t})} d\xi_t^{\Qx}
\end{align}
On the event $\tau>t$, we have $\Gamma(t,T;S_{t})={\Awesome \gamma(t;S_{t})}$ by definition of {\Awesome $\Gamma(t,T;S_t)$, and also 
$\gamma(t;S_{t}) = \left<\Pi(t;S_{t}), X_{t^-} \right>$}. Therefore, we can write the risk-neutral dynamics of prices process $\Gamma(t,T;S_t)$ as
\begin{align}\label{DRNM2call}
\frac{d\Gamma(t,T;S_t)}{\Gamma(t,T;S_{t})} = (r_t + h_t (L_t-1)) dt + \frac{{\left<{\Pi(t;{\Awesome S_{t}})}, dM^{\Qx}(t)\right>}}{{\left<{\Awesome \Pi(t;S_{t}), X_{t-}}\right>}}
+ {\Awesome \sigma_t} S_t {\Awesome \frac{\frac{\pa \left<\Pi(t;S_t), X_t \right>}{\pa s}}{\left<{\Awesome \Pi(t;S_{t}), X_{t^{-}}}\right>}} dW^{\Qx}_t - d\xi_t^{\Qx},\qquad t<\tau,
\end{align}
which concludes the second statement of the theorem.

\smallskip
\noindent
\textbf{(3)}
The price dynamics of the vulnerable claim under the real-world measure follows directly from plugging Eq. (\ref{RMM0}) into (\ref{DRNM2call}), and using the $\Px$ Brownian measure defined in Eq.~\eqref{eq:Qbrown} to get:
\begin{eqnarray}\label{DRNM2callhist}
\nonumber \frac{d\Gamma(t,T;S_t)}{\Gamma(t-,T;S_{t-})} &=& \bigg(r_t + h_t (L_t-1) + \frac{{\left<{\Pi(t;S_t)},(A'(t)-(A^{\mathbb{Q}})'(t))X_{t}  \right>}}{\left<{\Awesome \Pi(t;S_{t})}, X_{t-}\right>} + \frac{S_t {\Awesome \frac{\pa \left<\Pi(t;S_t),X_t \right>}{\pa s}}}{{\Awesome \left< \Pi(t;S_{t}), X_{t-} \right>}} \left(\mu_t - r_t \right) \bigg) dt \\
& & \quad +  \frac{{\left<{\Pi(t;S_t)}, dM^{\Px}(t)\right>}}{{\left<{\Awesome \Pi(t;S_{t}), X_{t-}}\right>}}
+ {\Awesome \sigma_t} S_t {\Awesome \frac{\frac{\pa \left<\Pi(t;S_t), X_t \right>}{\pa s}}{\left<{\Awesome \Pi(t;S_{t}), X_{t^{-}}}\right>}} dW_t  - d\xi_t^{\Px},\qquad t<\tau,
\end{eqnarray}
where we also used that $\xi^{\Px}=\xi^{\Qx}$ (see Lemma \ref{LmNdFD}).
Finally,  we get the dynamics (\ref{DRWP0call}) since
\begin{align}\label{AuxQnt3diff}
	 \frac{\left<\Pi(t;S_t),(A'(t)-(A^{\mathbb{Q}})'(t))X_{t}\right>}{\left<\Pi(t;S_t),X_{t}\right>}&=
	 \frac{\left<(A(t)-A^{\mathbb{Q}}(t))\Pi(t;S_t),X_{t}\right>}{\left<\Pi(t;S_t),X_{t}\right>}.
\end{align}
When $X_{t}=e_{i}$, the previous quantity simplifies as
\[
	 D_{i}(t,S_t):=\sum_{j=1}^{n}(a_{i,j}(t)-a_{i,j}^{\Qx}(t))\frac{\Pi_{j}(t,S_t)}{\Pi_{i}(t,S_t)},
\]
and (\ref{AuxQnt3diff}) can be written as $\left<(D_{1}(t,S_t),\dots,D_{N}(t,S_t))',X_{t}\right>$.}
\end{proof}

\begin{proof}[{\bf Proof of {\Blue Proposition} \ref{MnLmItrFrm}}]\hfill

\noindent
From the assumed time-invariance of the parameters, we have the following:
\[
\Phi_{i,m}[\Xi](\zeta) = \Ex_{i}^{\tilde{\mathbb{Q}}} \left[ \Xi(\tilde{X}_{m}) e^{-\sum_{n=0}^{m}{\zeta}(U_{(n+1)}-U_{(n)}) {\tilde{r}'}\tilde{X}_{n} - \sum_{n=1}^{m} \tilde{X}_{n-1}^{'}\tilde{K}\tilde{X}_{n}}  \right]
\]
Let $\Lambda_{n} := U_{(n)} - U_{(n-1)}$ for $n = 1, \dots, m+1$ and note that $\sum_{n=1}^{m+1} \Lambda_{n} = 1$.  It is {\Awesome well-known} \cite{Kendall} that the distribution of $$\Lambda:= (\Lambda_{1},\dots,\Lambda_{m})$$ is the symmetric Dirichlet distribution with parameter $\vec{\alpha} := (1,\dots,1) \in \mathbb{R}^{m+1}$. We recall that a random vector $\Lambda := (\Lambda_{1},\dots,\Lambda_{m})$ follows a Dirichlet distribution with parameters $\vec{\alpha} := (\alpha_{1},\dots,\alpha_{m+1}) \in \mathbb{R}^{m+1}$ such that $\min_{i} \alpha_{i} > 0$ if its density is given by
\[
	D(\lambda_{1},\dots,\lambda_{m})=\mathbf{B(\alpha)}^{-1}  \prod_{i=1}^{m} \lambda_{i}^{\alpha_{i}-1}(1-\lambda_{1}- \dots - \lambda_{m})^{\alpha_{m+1}-1}{\bf 1}_{T_{m}}(\lambda_{1},\dots,\lambda_{m}),
\]
where
$\mathbf{B(\alpha)} := \prod_{i=1}^{m+1} \Gamma (\alpha_{i}) /\Gamma \left(\sum_{i=1}^{m+1}\alpha_{i} \right)$ and $ T_{m}:= \{(\lambda_{1},\dots,\lambda_{m}) \in \mathbb{R}^{m} : \lambda_{i} \geq 0 \; \& \; \sum_{i=1}^{m}\lambda_{i} \leq 1 \}$.
Next, conditioning on $(\widetilde{X}_{1},\dots,\widetilde{X}_{m})$ and using independence between $U$ and $\tilde{X}$,
\begin{equation*}
 \Phi_{i,m}[\Xi](\zeta) = \Ex_{i}^{\tilde{\mathbb{Q}}} \Big[\int_{T_{m}}e^{-\sum_{n=0}^{m}\zeta{\tilde{r}'}\tilde{X}_{n}\lambda_{n+1}} D(\lambda_{1},\dots,\lambda_{m})d\lambda_{1}\dots d\lambda_{m} \Xi(\tilde{X}_{m})e^{-\sum_{n=1}^{m}\tilde{X}_{n-1}^{'}\tilde{K}\tilde{X}_{n}} \Big],
\end{equation*}
where $\lambda_{m+1}=1-\sum_{{\DGold{n=1}}}^{m}\lambda_{n}$. The following simplification can be made:
\begin{align*}
\int_{T_{m}}e^{-\sum_{n=0}^{m}\zeta{\tilde{r}'}\tilde{X}_{n}\lambda_{n+1}} D(\lambda_{1},\dots,\lambda_{m}) d\lambda_{1}\dots d\lambda_{m}&= \int_{T_{m}}e^{-\sum_{n=0}^{m-1} \zeta{\tilde{r}'} \left(\tilde{X}_{n} - {\tilde{X}_{m}} \right) \lambda_{n+1} -\zeta {\tilde{r}'}\tilde{X}_{m}}  D(\lambda_{1},\dots,\lambda_{m}) d\lambda_{1}\dots d\lambda_{m}\\
&= e^{-\zeta {\tilde{r}'} \tilde{X}_{m}} \mathcal{L}_{m} \left( \zeta {\tilde{r}'}( \tilde{X}_{0} - \tilde{X}_{m} ), \dots, \zeta {\tilde{r}'}(\tilde{X}_{m-1} - \tilde{X}_{m}) \right),
\end{align*}
where $\mathcal{L}_{m}$ is given as in (\ref{DrcLap}). Finally, note that, by construction, $\tilde\Qx(\tilde{X}_{i}=\tilde{e}_{k}|\tilde{X}_{i-1}=\tilde{e}_{m})=1/(N-1)$ for any $k\neq{}m$ and, thus,
\[
	\tilde{\Qx}(\tilde{X}_{1}=\tilde{e}_{1},\dots,\tilde{X}_{m} =\tilde{e}_{m}| \tilde{X}_{0}=e_{i}) =  \frac{1}{(N-1)^{m}},
\]
for all $(\tilde{e}_{1},\dots,\tilde{e}_{m})\in\{e_{1},\dots,e_{N}\}^{m}$ such that $\tilde{e}_{i}\neq\tilde{e}_{i-1}$. In that case, it is clear that
\begin{equation*}
{\Phi_{i,m}[\Xi](\zeta) = \displaystyle{\sum_{ \overset{\left( \tilde{e}_{1},\dots,\tilde{e}_{m} \right)}{e_{i} \neq e_{i+1}}}} \Xi(\tilde{e}_{m})e^{-\zeta {\tilde{r}'}\tilde{e}_{m} {-} \sum_{n=1}^{m} \tilde{e}_{n-1}^{'} \tilde{K}\tilde{e}_{n} } \mathcal{L}_{m} \left( \zeta{\tilde{r}'}( \tilde{e}_{0} - \tilde{e}_{m} ), \dots, \zeta{\tilde{r}'}(\tilde{e}_{m-1} - \tilde{e}_{m}) \right) \frac{1}{(N-1)^{m}}.}
\end{equation*}
\end{proof}

\begin{proof}[{\bf Proof of {\Blue Theorem} \ref{Lem:MnEstErr}}]\hfill

\noindent
{\Blue Let  ${\bf 1}(e_{i})=1$ for all $i=1,\dots,N$, and   $\|\Xi\|_{\infty}=\max_{\ell} |\Xi(e_{\ell})|$.
Let us first note that for all $u\leq v$ with $v-u\leq \delta$,
\begin{align}\label{Eq:EasyEstForNrm}
	\left|\widetilde{\Psi}_{i}^{(M,r)}[\Xi](u;v)\right|\leq
	\|\Xi\|_{\infty}
	\sum_{m=0}^{M-1} e^{-(v-u)}\frac{(v-u)^{m}}{m!}\Phi_{i,m}[{\bf 1}](v-u;v)\leq \|\Xi\|_{\infty} \Psi_{i}[{\bf 1}](v-u;v)\leq \|\Xi\|_{\infty},
\end{align}
where for the second inequality above we have use (\ref{Eq:ApprxPhiCond2}) and the fact that $\Psi_{i,m}[{\bf 1}]\leq 1$ (being the bond price under the $i^{th}$ economic regime). {\Blue In particular, from the definition (\ref{Eq:BkwrdDfnPsiApp}),
\begin{align*}
	 \max_{i}\left|\widetilde{\Psi}^{(M,r,k)}_{i}[{\Xi}](j\delta;T)\right|&=\max_{i}\left|\widetilde{\Psi}^{(M,r)}_{i}[\widetilde{\Xi}^{(j)}](j\delta;(j+1)\delta)\right|\\
	&\leq
	\max_{i}\left|\widetilde{\Xi}^{(j)}(e_{i})\right|= \max_{i}\left|\widetilde{\Psi}^{(M,r,k)}_{i}[{\Xi}]((j+1)\delta;T)\right|\\
	&\leq \dots \leq  \max_{i}\left|\widetilde{\Psi}^{(M,r,k)}_{i}[{\Xi}]((k-1)\delta;k\delta)\right|\leq \|\Xi\|_{\infty}
\end{align*}
Thus, by backward induction,} it follows that
\begin{equation}\label{Eq:UBIOp}
	\max_{j=0,\dots,K-1}\max_{i=1,\dots,N}\left|\widetilde{\Psi}^{(M,k,r)}_{i}[\Xi](j \delta;T)\right|\leq \|\Xi\|_{\infty}
\end{equation}
{\Blue Next, note that,} for $j=k-1$ and {\Blue $D:=\|\Xi\|_{\infty}(1+B)$},
\begin{align}
	\left|\widetilde{\Psi}^{(M,k,r)}_{i}[\Xi](j \delta;T)-{\Psi}_{i}[\Xi](j \delta;T)\right|&=\left|\widetilde{\Psi}^{(M,r)}_{i}[\Xi](j \delta;T)-{\Psi}_{i}[\Xi](j \delta;T)\right|\nonumber \\
	&\leq{}
	\sum_{m=r}^{M-1}e^{-\delta}\frac{\delta^{m}}{m!}\left|\widetilde{\Phi}_{i,m}[\Xi](\delta;T)-{\Phi}_{i,m}[\Xi](\delta;T)\right| +\sum_{m=M}^{\infty}e^{-\delta}\frac{\delta^{m}}{m!}\left|{\Phi}_{i,m}[\Xi](\delta;T)\right| \nonumber  \\
	&\leq B {\Blue \|\Xi\|_{\infty}} \sum_{m=r}^{M-1}e^{-\delta}\frac{\delta^{m+j}}{m!}+\|\Xi\|_{\infty}\sum_{m=M}^{\infty}e^{-\delta}\frac{\delta^{m}}{m!}{\Phi}_{i,m}[{\bf 1}](\delta;T)\nonumber \\
	&\leq D \delta^{(r+j)\wedge M}, \label{Eq:NDvl1}
\end{align}
where we used the representation (\ref{CmpPsii}), (\ref{Eq:ApprxPhiCond}), and the fact that
\[
	\sum_{m=M}^{\infty}e^{-\delta}\frac{\delta^{m}}{m!}{\Phi}_{i,m}[{\bf 1}](\delta;T)\leq \delta^{M}\sum_{\ell=0}^{\infty}e^{-\delta}\frac{\delta^{\ell}}{\ell!}{\Phi}_{i,\ell}[{\bf 1}](\delta;T)=\delta^{M}\Psi_{i}[{\bf 1}](T-\delta;T)\leq{}\delta^{M}.
\]
Next, for $j=k-2$, using (\ref{SmallTimeExp2}) {\Blue and  (\ref{Eq:BkwrdDfnPsiApp})},
\begin{align}
	\left|\widetilde{\Psi}^{(M,k,r)}_{i}[\Xi](j \delta;T)-{\Psi}_{i}[\Xi](j \delta;T)\right|&\leq \underbrace{\left| \widetilde{\Psi}^{(M,r)}_{i}[\widetilde{\Xi}^{(j)}](j\delta;(j+1)\delta) -
	{\Psi}_{i}[\widetilde{\Xi}^{(j)}](j\delta;(j+1)\delta)
	\right|}_{T_{1}} \label{DT1Nd}\\
	&+ \underbrace{\left|{\Psi}_{i}[\widetilde{\Xi}^{(j)}](j\delta;(j+1)\delta) -
	{\Psi}_{i}[\widehat{\Xi}^{(j)}](j\delta;(j+1)\delta)
	\right|}_{T_{2}}, \label{DT2Nd}
\end{align}
where
\begin{align*}
\widetilde{\Xi}^{(j)}(e_{i}):= \widetilde{\Psi}^{(M,r,k)}_{i}[{\Xi}]((j+1)\delta;T),
\qquad
\widehat{\Xi}^{(j)}(e_{i}):= {\Psi}_{i}[{\Xi}]((j+1)\delta;T).
\end{align*}
The term in (\ref{DT1Nd}) can be bounded analogously as in (\ref{Eq:NDvl1}){\Blue ,} but replacing the constant $D$ by {\Blue $\|\widetilde{\Xi}^{(j)}\|_{\infty}(1+B)$}, which itself can be bounded {\Blue again by $D=\|{\Xi}\|_{\infty}(1+B)$} in light of (\ref{Eq:UBIOp}). Thus,
\[
	T_{1}\leq D \delta^{(r+j)\wedge M}.
\]
Similarly, we have
\[
	T_{2}=\left|{\Psi}_{i}[\widetilde{\Xi}^{(j)}-\widehat{\Xi}^{(j)}](j\delta;(j+1)\delta) \right|\leq \max_{\ell}\left|\widetilde{\Xi}^{(j)}(e_{\ell})-\widehat{\Xi}^{(j)}(e_{\ell})\right|
	{\Psi}_{i}[{\bf 1}](j\delta;(j+1)\delta)\leq \max_{\ell}\left|\widetilde{\Xi}^{(j)}(e_{\ell})-\widehat{\Xi}^{(j)}(e_{\ell})\right|,
\]
which can be bounded by $D \delta^{(r+j)\wedge M}$ in light of (\ref{Eq:NDvl1}). Putting together the previous two estimates,
\[
	\left|\widetilde{\Psi}^{(M,k,r)}_{i}[\Xi]({\Blue (k-2)} \delta;T)-{\Psi}_{i}[\Xi]({\Blue (k-2)\delta};T)\right|\leq {\Blue D \delta^{(r+j)\wedge M}}.
\]
Proceeding by induction, it follows that
\[
	{\Blue \left|\widetilde{\Psi}^{(M,k,r)}_{i}[\Xi]({\Blue \ell} \delta;T)-{\Psi}_{i}[\Xi]({\Blue \ell} \delta;T)\right|\leq (k-{\Blue \ell}) D \delta^{(r+j){\Blue \wedge} M}=(k-\ell)D\left(\frac{T}{k}\right)^{(r+j)\wedge M},}
\]
{\Blue and, in particular, we obtain (\ref{Eq:KIFEAn}) by setting $\ell=0$}.}
\end{proof}

\begin{proof}[{\bf Proof of Corollary \ref{Lem:MnEstErrMod}}]\hfill

\noindent
{\Blue The proof is very similar to that of {\DViolet Theorem} \ref{Lem:MnEstErr}. The key difference is in the estimate (\ref{Eq:EasyEstForNrm}), which needs to be modified. Under the modified condition {\Blue (\ref{Eq:ApprxPhiCond2Milder})}, we proceed {\DRed as} follows, fixing $\zeta=v-u$,
\begin{align*}
	\left|\widetilde{\Psi}_{i}^{(M,r)}[\Xi](u;v)\right|&=\bigg|\sum_{m=0}^{r-1} e^{-\zeta}\frac{\zeta^{m}}{m!}\Phi_{i,m}[\Xi](\zeta;v)+
		\sum_{m=r}^{M-1} e^{-\zeta}\frac{\zeta^{m}}{m!}\widetilde{\Phi}_{i,m}[\Xi](\zeta;v)\bigg|\\
	&\leq \|\Xi\|_{\infty}
	\sum_{m=0}^{\infty} e^{-\zeta}\frac{\zeta^{m}}{m!}\left|\Phi_{i,m}[{\bf 1}](\zeta;v)\right|+ \|\Xi\|_{\infty} C
		\sum_{m=r}^{M-1} e^{-\zeta}\frac{\zeta^{m}}{m!}\\
	&\leq \|\Xi\|_{\infty}\Psi_{i}[{\bf 1}](\zeta;v)+ \|\Xi\|_{\infty} C \zeta^{r}\\
	&\leq \|\Xi\|_{\infty}\left(1+ C \zeta^{r}\right).
\end{align*}
In that case, by backward induction,  Eq.(\ref{Eq:UBIOp}) will take the form:
\begin{equation}\label{Eq:UBIOpMod}
	\max_{j=0,\dots,K-1}\max_{i=1,\dots,N}\left|\widetilde{\Psi}^{(M,k,r)}_{i}[\Xi](j \delta;T)\right|\leq \|\Xi\|_{\infty}
	\left(1+C\delta^{r}\right)^{k}=  \|\Xi\|_{\infty}
	\left(1+C\frac{T^{r}}{k^{r}}\right)^{k}\leq F \|\Xi\|_{\infty},
\end{equation}
for a constant $1\leq F<\infty$ (depending only on $C$, $T$, and $r$). Following the same proof as in {\DViolet Theorem} \ref{Lem:MnEstErr}, we will have, by induction, that
\[
	{\Blue \left|\widetilde{\Psi}^{(M,k,r)}_{i}[\Xi]({\Blue \ell} \delta;T)-{\Psi}_{i}[\Xi]({\Blue \ell} \delta;T)\right|\leq (k-{\Blue \ell}) E \delta^{(r+j){\Blue \wedge} M}=(k-\ell)E\left(\frac{T}{k}\right)^{(r+j)\wedge M},}
\]
with $E:=\|\Xi\|_{\infty} (B+ F)$ and Eq. (\ref{Eq:KIFEAnMod}) follows.}
\end{proof}

\section{Pseudo-codes}\label{Sec:PseudoCodesAll}

\begin{algorithm}[h!]
\caption{[Prices] = PriceClaim($T$, $\Xi$, $\delta$, $M$)}
\floatname{algorithm}{Procedure}
\label{alg:FindOptPrice}
\begin{algorithmic}
\STATE  $\zeta= \min\{T, \delta\}$
\STATE\textbf{for} $i=1,\dots,N$
\STATE     \hspace{0.4 cm} ${\rm Prices}(i,1) = e^{-\zeta}\sum_{m=0}^{M-1}\frac{\zeta^{m}}{m!}\,\Phi_{i,m}[\Xi](\zeta)$
\STATE \textbf{endfor}
\STATE  \textbf{if} $T\leq \delta$
\STATE     \hspace{0.4 cm} \textbf{Return} Prices(:,1)
\STATE \textbf{else}
\STATE \hspace{0.4 cm} \textbf{for} $j = 2, 3, \dots, \lceil \frac{T}{\delta}\rceil $
\STATE \hspace{0.8 cm} \textbf{for} $i=1,\dots,N$
\STATE     \hspace{1.2 cm} ${\rm Prices}(i,j) = e^{-\delta}\sum_{m=0}^{M-1}\frac{\delta^{m}}{m!}\,\Phi_{i,m}[\rm{Prices(:,j-1)}](\delta)$
\STATE  \hspace{0.8 cm} \textbf{endfor}
\STATE  \hspace{0.4 cm} \textbf{endfor}
\STATE \textbf{endif}
\STATE  \textbf{Return} ${\rm Prices}(:,:)$
\end{algorithmic}
\end{algorithm}

\begin{algorithm}[h!]
\caption{[Price] = ComputeVulnOption($T$, $\delta$, $\Delta$,$M$,k)}
\floatname{algorithm}{Procedure}
\label{alg:FindVulnOptPrice}
\begin{algorithmic}
\STATE \textbf{if} $T\leq{} \delta$
\STATE \hspace{0.4 cm}\textbf{for} $i=-B,-B+1,\dots,B$
\STATE \hspace{0.8 cm}\textbf{for} $j=1, 2,\dots,N,\dots,B$
\STATE     \hspace{1.2 cm}$\Xi_{i}(j) ={\rm BS}\left(T;se^{i \Delta+T r_{j}},\sigma_{j}^{2},0,K\right)$
\STATE \hspace{0.8 cm}\textbf{endfor}
\STATE     \hspace{0.8 cm}$[{\rm Price}(i,1),\dots,{\rm Price}(i,N)]= {\rm PriceClaim}(T,\Xi_{i},T/k,M)$
\STATE \hspace{0.4 cm}\textbf{endfor}
\STATE \textbf{else}
\STATE     \hspace{0.4 cm}${\rm Price}_{new}=\text{ComputeVulnOption}(T-\delta, \delta, \Delta, M, k)$
\STATE \hspace{0.4 cm}\textbf{for} $i=-B,-B+1,\dots,B$
\STATE \hspace{0.8 cm}\textbf{for} $j=1, 2,\dots,N,\dots,B$
\STATE     \hspace{1.2 cm}$\widetilde{\Xi}_{i}(j) =\sum_{k=-B}^{B} {\rm Price}_{new}(k,j)\int_{z_{k-1}^{i,j}}^{z_{k}^{i,j}} \frac{1}{\sqrt{2\pi}}e^{-z^{2}/2}dz$
\STATE \hspace{0.8 cm}\textbf{endfor}
\STATE     \hspace{0.8 cm}$[{\rm Price}(i,1),\dots,{\rm Price}(i,N)]= {\rm PriceClaim}(\delta, \widetilde{\Xi}_{i}, \delta/k, M)$
\STATE \hspace{0.4 cm}\textbf{endfor}
\STATE \textbf{endif}
\end{algorithmic}
\end{algorithm}

\section{Computational speed comparisons}\label{Sec:SpeedCmp}

Table \ref{tab:SpeedComp3meth} compares the computational times for our proposed method (Algorithm \ref{alg:FindOptPrice}) with respect to both the Runge-Kutta and a matrix exponential approximation method based on the representation (\ref{eq:psiexpr}) using the parameter setup of Table \ref{tab:histgen} given {in} Section \ref{sec:simscenario}.  For each given pair $(\delta,T)$, the bond prices $(\psi_{1}(\zeta),\psi_{2}(\zeta),\psi_{3}(\zeta))$ {are} computed for all time-to-maturity $\zeta\in \{\delta, 2\delta, \dots, k\delta := T\}$.  The entries in Table \ref{tab:SpeedComp3meth} represent the ratio of the new algorithm's computational time with $M=2$ to the respective ODE Method computational time and the matrix exponential computational time.  As we can see, our method tends to outperform  ODE Method for maturities shorter than 15 years and also for mesh sizes larger than $0.05$ years; however, even in the worst case our method operates at a speed which is very comparable to the faster of the two.

\begin{table}[ht]
\begin{center}
\begin{tabular}{|c|c|c|c|c|c|c|c|c|}
\hline
\multicolumn{9}{|c|}{\textbf{Relative comp. times of new meth. vs. Runge-Kutta (N=3)}} \\
 \hline
\multicolumn{1}{|c|}{} & \multicolumn{8}{|c|}{Time Horizons, T} \\
  \hline
$\delta$& 5 & 10 & 15 & 20 & 25 & 30 & 35 & 40 \\
  \hline
0.1   & 0.3093 & 0.4102 & 0.5039 & 0.5706 & 0.6701 & 0.5112 & 0.6227 & 0.7867 \\
0.075 & 0.2990 & 0.3912 & 0.4245 & 0.5749 & 0.7545 & 0.9124 & 1.1011 & 1.2094 \\
0.05  & 0.3655 & 0.4212 & 0.5828 & 0.7571 & 0.9248 & 0.8193 & 1.1855 & 1.1843 \\
0.025 & 0.4095 & 0.7270 & 1.0501 & 1.1125 & 1.5078 & 1.4490 & 1.6683 & 1.8545 \\
\hline
\multicolumn{9}{|c|}{\textbf{Relative comp. times of new meth. vs. matrix exp. (N=3)}} \\
\hline
\multicolumn{1}{|c|}{} & \multicolumn{8}{|c|}{Time Horizons, T} \\
\hline
$\delta$ & 5 & 10 & 15 & 20 & 25 & 30 & 35 & 40 \\
\hline
0.1 & 0.3828 & 0.4121 & 0.4490 & 0.5634 & 0.6183 & 0.5545 & 0.6515 & 0.7722 \\
0.075 & 0.3727 & 0.5049 & 0.6791 & 0.8080 & 0.6502 & 0.8990 & 0.8664 & 0.9054 \\
0.05 & 0.4149 & 0.7151 & 0.5571 & 0.7468 & 0.9756 & 1.0217 & 0.9247 & 1.1846 \\
0.025 & 0.5757 & 0.6770 & 0.9240 & 1.1613 & 1.0855 & 1.2903 & 1.4646 & 1.1325 \\
\hline
\end{tabular}
\caption{{The top panel shows the ratio of the processing time using our proposed method (Algorithm \ref{alg:FindOptPrice} above) and a Runge-Kutta Type numerical solution of (\ref{eq:bondspdeGen}) using the MATLAB function ``ode45". The bottom panel shows the processing time ratio between our method and a Pad\'e type approximation of the matrix exponential (\ref{eq:psiexpr}) using the MATLAB command ``expm''.}}
\label{tab:SpeedComp3meth}
\end{center}
\end{table}


\end{document}